\documentclass[%
 reprint,
 amsmath,amssymb,
 aps,
]{revtex4-2}

\usepackage{graphicx}% Include figure files
\usepackage{dcolumn}% Align table columns on decimal point
\usepackage{bm}% bold math
\usepackage{hyperref}% add hypertext capabilities
\usepackage{subfigure}
\graphicspath{{figures/}}
\usepackage[table]{xcolor}

\usepackage{enumitem}

\newcommand{\etal}{\textit{et al.~}}

\begin{document}

\preprint{APS/123-QED}
\maxdeadcycles=500

\title{A fractional stochastic theory for interfacial polarization of cell aggregates}

\author{Pouria A. Mistani$^1$}
 \altaffiliation[Corresponding author: ]{pouria@ucsb.edu}
\author{Samira Pakravan$^1$}
\author{Frederic Gibou$^{1,2,3}$}
\affiliation{
 $^1$ Department of Mechanical Engineering,  $^2$ Department of Computer Science, $^3$ Department of Mathematics, University of California Santa Barbara, Santa Barbara, CA 93106, USA\\
}

\date{\today}% It is always \today, today,

\begin{abstract}
We present a theoretical framework to model the electric response of cell aggregates. We establish a coarse representation for each cell as a combination of membrane and cytoplasm dipole moments. Then we compute the effective conductivity of the resulting system, and thereafter derive a Fokker-Planck partial differential equation that captures the time-dependent evolution of the distribution of induced cellular polarizations in an ensemble of cells. Our model predicts that the polarization density parallel to an applied pulse follows a skewed t-distribution, while the transverse polarization density follows a symmetric t-distribution, which are in accordance with our direct numerical simulations. Furthermore, we report a reduced order model described by a coupled pair of ordinary differential equations that reproduces the average and the variance of induced dipole moments in the aggregate. We extend our proposed formulation by considering fractional order time derivatives that we find necessary to explain anomalous relaxation phenomena observed in experiments as well as our direct numerical simulations. Owing to its time-domain formulation, our framework can be easily used to consider nonlinear membrane effects or intercellular couplings that arise in several scientific, medical and technological applications.
\end{abstract}

%Use showkeys class option if keyword
\maketitle

%\tableofcontents

%\frederic{Let's make sure that we only number the equations that we refer to or number them all.}

%\frederic{Let's make sure that all the variables are defined, even the obvious one $\epsilon_0$, $R_1$, etc.}

\section{Introduction}
Effects of external electric fields on heterogeneous systems have been of great scientific and technological importance throughout the past century. These systems include composite materials, colloidal suspensions, and biological cells. In the case of biological cells, electric fields have found several applications for cell fusion \cite{zimmermann1982electric,jordan2013electroporation}, electrorotation \cite{arnold1984electric,fuhr1986rotation}, dielectrophoresis \cite{pohl1971dielectrophoresis,sauer1983forces} and cancer cell separation \cite{becker1995separation,gascoyne2002particle}, electroporation \cite{neumann1972permeability}, levitation \cite{kaler1990dielectrophoretic} and cell deformation \cite{bryant1987electromechanical}. For an early review of its applications in biotechnology and medicine we refer to Markx and Davey (1999) \cite{markx1999dielectric}. More recently, transmembrane potential (TMP) patterns that emerge in multicellular living organisms have gained extra attention due to discovery of their regulatory role in development and regeneration; we refer to the review of Levin \textit{et al.} (2017) for a comprehensive overview on developmental bioelectricity \cite{levin2017endogenous}.  Even though modern research on TMP manipulations are focused on molecular based treatments, it has been long known that TMP patterns altered by external electric fields could influence development \cite{roux1uber}, embryogenesis \cite{hotary1992evidence,pullar2016physiology} or wound healing \cite{reid2014electrical}. Therefore, developing a predictive and generalizable mathematical model to understand the effects of different cellular mechanisms on the tissue level bioelectric patterns poses promising opportunities in bioengineering.

In all of these applications, it is essential to have a precise knowledge of the TMP induced over cell membranes, especially in cell aggregates that are composed of tens of thousands of cells with a heterogeneous mix of morphologies and electrical properties. Even though much is known about single cell electric interactions, a theory that predicts a detailed distribution of transmembrane potentials within cell aggregates has been missing. In this work, we present a novel theoretical framework based on a Fokker-Planck description, for tracing the time evolution of the probability density of multicellular polarizations in response to arbitrary electric stimulations. We further introduce a moment-based analytic reduced-order model of the proposed Fokker-Planck equation that provides statistics of transmembrane potentials with minimal computational expense. Importantly, besides multicellular systems, our theory is applicable to a broad range of systems such as shelled colloidal particles, emulsions and composite materials \cite{asami2002characterization}. Moreover, it can be extended to include the effects of membrane nonlinearities \cite{mistani2020prep} or intercellular couplings \cite{cervera2020bioelectrical}, counterion polarizations, and eventually real-time pulse optimization, which is of great benefit to emerging biomedical treatments using electric fields such as electrochemotherapy.

\subsection{Physical bioelectric processes}\label{Physical_bioelectric_processes}
Electrical properties of biological tissues have been extensively investigated for the last two centuries since the discovery of Ohm's law. A comprehensive historical overview of different aspects of biological dielectric response (including basic concepts, tabulated data, underlying molecular processes and effective cellular interactions) is covered in the surveys of Schwan (1957) \cite{schwan1957electrical}, Stuchly and Stuchly (1980) \cite{stuchly1980dielectric}, Pethig (1984) \cite{pethig1984dielectric}, Pethig and Kell (1987) \cite{pethig1987passive}, Foster and Schwan (1989) \cite{foster1986dielectric}, McAdams and Jossinet (1995) \cite{mcadams1995tissue}, Gabriel (1996) \cite{gabriel1996dielectric}, Kuang and Nelson (1998) \cite{kuang1998low}, with more recent reviews on its different applications such as electroporation provided by Kotnik \textit{et al.} (2019) \cite{kotnik2019membrane}. 

Early research on the electric response of bulk biological materials revealed that tissues exhibit resistive and capacitive behaviors. Experiments showed an early peak in current in response to a step voltage, which could be attributed to an increasing tissue resistance, or from an induced counter-potential polarization. Furthermore, with the advent of high frequency apparatus it was possible to examine the high frequency response of tissues, which led to the recognition that tissues exhibit low resistance at higher frequencies and the dielectric response of tissues was determined by different physical processes at each frequency regime. Bulk electric properties are mainly determined by cell membranes and cellular structures. At the cellular level, three main physical processes, namely interfacial polarization, ionic diffusion and dipolar orientation of polar molecules, are identified to play key roles in dielectric dispersions at different frequency regimes \cite{schwan1957electrical}. While the origin of $\alpha$- and $\gamma$-dispersions are relaxation processes in the bulk phases of the material, $\beta$-dispersions originate from internal boundary conditions imposed by interfaces separating different phases; this is the focus of the current work. Below, we briefly review these mechanisms.

\textbullet\ $\alpha$-dispersion: The main factor that contributes to the \textit{$\alpha$-dispersion} (at audio or sub-KHz frequencies) is the ionic diffusion in the electric double layer in the immediate vicinity of charged surfaces as well as in the bulk. $\alpha$-dispersion is characterized with high dielectric constants at low frequencies. Schwan first observed this mode of polarization at low frequencies in biological tissues \cite{schwan1957electrical} and later, Schwan and co-workers showed that this effect is also observed in non-biological colloids \cite{schwan1962low}. Schwarz (1962) \cite{schwarz1962theory} was the first to develop a theory that took into account the counterion polarization around colloidal particles suspended within electrolytes. Schwarz showed these displacements could be modeled by an additional ``apparent'' dielectric constant that reach high values at low frequencies. Schwarz's method did not consider diffusion in the double layer itself, and was later extended by Einolf and Carstensen (1971) \cite{einolf1971low,einolf1973passive} to include diffusion on both sides of the membranes. Dukhin and Shilov (1974) \cite{dukhin1974dielectric} proposed a more accurate treatment by considering ionic diffusion in the bulk, rather than just the thin layers around charged particles (see also the review by Mandel and Odijk \cite{mandel1984dielectric}). A simplified formulation of Dukhin's model that admits analytical solution was given by Grosse and Foster \cite{grosse1987permittivity}, which helped to show that the corresponding Cole-Cole plot is broader than the Debye's model, indicating that counter-ion polarization is partly responsible for the observed anomalous relaxation of biological matter. In short, counter-ion polarization theories explain $\alpha$-dispersion and predict high permittivities at low frequencies that exhibit broad relaxation behaviors.

\textbullet\ $\beta$-dispersion: Interfacial polarization dominates the dielectric properties of tissues at \textit{$\beta$-dispersion} (radio frequencies from tens of KHz to tens of MHz range, timescales determined by membrane resistance and capacitance) as well as the dielectric properties of colloids and emulsions. Biological mixtures of interest to us have a \textit{triphasic} dielectric structure with conductive parts composed of a cytoplasm covered by a membrane immersed in a continuous medium. Historically, dielectric theories of interfacial polarization began by considering \textit{diphasic suspensions} in the seminal treatise of Maxwell (1873) \cite{maxwell1873treatise} and later Wagner (1914) \cite{wagner1914erklarung}. In 1925, Fricke \cite{fricke1925mathematical} developed a dielectric theory for spherical particles surrounded by nonconductive membranes and extended it to membrane-covered ellipsoidal particles in 1953 \cite{fricke1953electric}. Maxwell-Wagner theory has the following limitations: (i) it is only valid at very low concentrations and assumed that the local electric field was equal to the external electric field, (ii) the interior of particles were assumed to be at constant potential, and (iii) the external field was modeled as if the particle was a perfect insulator. Hanai (1960) \cite{hanai1960theory} developed an interfacial polarization theory that was valid at high concentrations, by assuming Wagner's relation holds during successive additions of infinitesimally small quantities of the disperse phase. Hanai and co-workers (1979) \cite{hanai1979dielectric} later generalized their approach to the case of suspensions of shelled spheres and Zhang \textit{et al.} (1983-1984) \cite{zhang1983dielectric,zhang1984dielectric} showed that the theory could explain experiments with polystyrene microcapsules. This strategy was applied to three-phase structures in 1993 \cite{hanai1993theoretical}. We shall emphasize that even though Hanai's approach is advantageous over Wagner's theory as it holds its validity to high concentrations, it is still based on the non-conductive assumption for membranes and, more importantly, it is not clear how to consider nonlinear variations in the membrane conductance during the application of electric pulses similar to the case of electroporation. Our theory builds on this line of work and aims to address these shortcomings by constructing a time-domain model for interfacial polarization in cell aggregates. Unlike the aforementioned works that are limited to modeling average properties at the aggregate level, our approach captures detailed information about the distribution of induced polarizations as well as their time-dependent evolution. 

\textbullet\ $\gamma$-dispersion: The third mechanism that is responsible for the \textit{$\gamma$-dispersion} (microwave frequencies from MHz to GHz range) is the dipolar orientations of permanent polar molecules, \textit{e.g.} water molecules and other macromolecules. Under an applied electric torque and opposed by thermal agitations in the medium, polar molecules undergo rapid reorientations towards thermal equilibrium and exhibit dielectric relaxation. This phenomenon is described by Debye's theory (1929) \cite{debye1929polar},  which is inherently a Fokker-Planck equation that describes the evolution of the probability density of dipolar orientations under an applied pulse. Our theory presented here is inspired by this strategy.

\subsection{Equations of interfacial polarization}
In its general form, Maxwell's equations in matter read
\begin{align}
&\nabla\cdot \mathbf{D}=\rho_f   \label{eq::MX1}  \\
&\nabla\cdot \mathbf{B}=0 \label{eq::MX2}\\
&\nabla\times \mathbf{E}=-\partial_t \mathbf{B}   \label{eq::MX3}\\
&\nabla\times \mathbf{H}=  \mathbf{J}_f + \partial_t \mathbf{D}\label{eq::MX4}
\end{align}
where $\rm E$ and $\rm B$ are the electric and magnetic fields, respectively, $\rm \mathbf{D}=\epsilon \mathbf{E}$, $\rm \mathbf{H}=\mu\mathbf{B}$ and the total current is defined by $\rm \mathbf{J}=\mathbf{J}_f + \partial_t \mathbf{D}$. However, electric interactions within a multicellular system can be modeled under the \textit{quasi-electrostatic} assumption, \textit{i.e.} when the wavelength of the stimulating electric pulse is larger than the cell size. Under the quasi-electrostatic assumption, the induced magnetic fields are negligible and therefore the electric field is curl free and we may define the electric potential $\rm u$ by the relation $\rm \mathbf{E}=-\nabla u$. Also, computing the divergence of \eqref{eq::MX4} and using \eqref{eq::MX1}, we have that $\rm \nabla \cdot (\sigma \nabla u)=\dfrac{\partial \rho_f}{\partial t}$, where in the absence of a net free charge density $\rm \rho_f$, when only interfacial polarization is present, the electric field is given by the solution of the Laplace equation $\rm \nabla \cdot (\sigma\nabla u)=0$, and we can neglect the permittivity of the cytoplasm and  of the extra-cellular medium.
\begin{figure}
\begin{center}
\subfigure{\includegraphics[width=\linewidth]{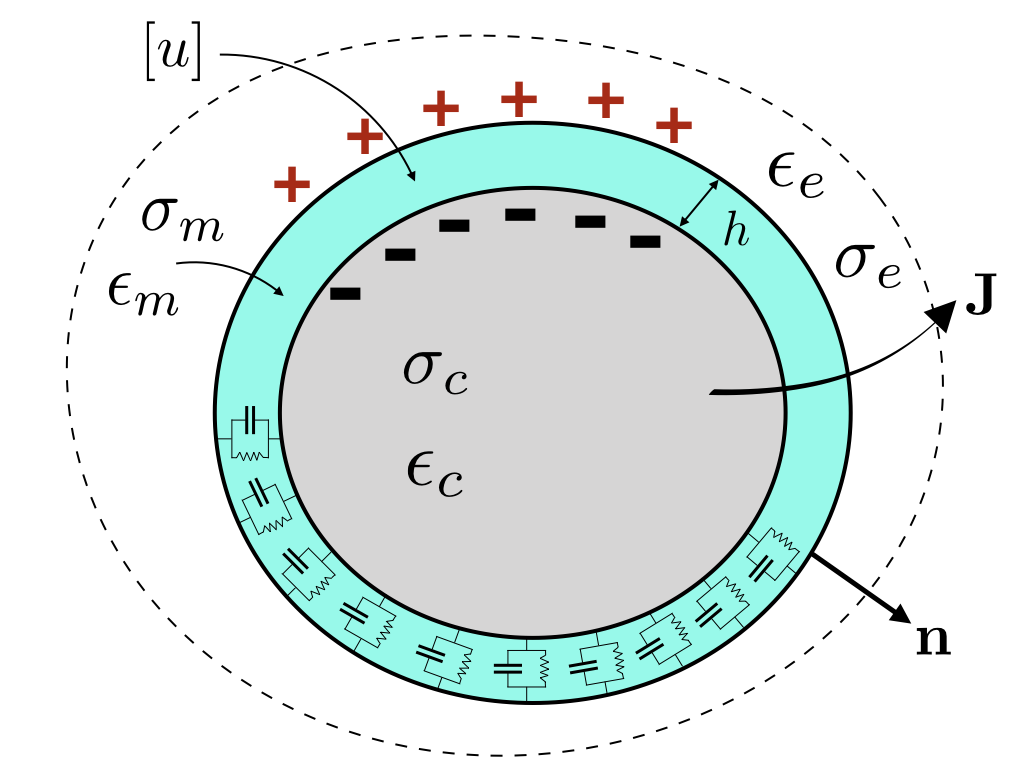} } \quad \quad
\end{center}
\caption{Cell membranes are modeled with an array of parallel resistors and capacitors in a thin layer surrounding the cytoplasm. Here we consider a sharp interface.}
\label{fig::bilayer}
\end{figure} 

As for cells, we consider a shelled particle model with two concentric surfaces $\rm \Gamma^\pm$ forming a membrane with thickness $\rm h$ (see figure \ref{fig::bilayer}). The boundary conditions impose that the electric potential must be continuous across the interfaces
\begin{align*}
&u_c(\mathbf{x}^-)=u_m(\mathbf{x}^-), &u_m(\mathbf{x}^+)=u_e(\mathbf{x}^+),
\end{align*}
with $\rm \mathbf{x}^\pm\in \Gamma^\pm$ and that the normal component of the total current $\rm \mathbf{J}_k=(\sigma_k + j\omega\epsilon_0 \epsilon_k)\mathbf{E}_k=\Lambda_k^\ast \mathbf{E}_k$ (with $\rm k=e, m$ and $\rm c$ referring to the extra-cellular medium, the membrane and the cytoplasm, respectively) must be continuous across the boundaries:
\begin{align*}
&\Lambda_c^\ast \partial_n u_c (\mathbf{x})=\Lambda_m^\ast \partial_n u_m (\mathbf{x}), &\mathbf{x}\in \Gamma^-,\\
&\Lambda_m^\ast \partial_n u_m (\mathbf{x})=\Lambda_e^\ast \partial_n u_e (\mathbf{x}), &\mathbf{x}\in \Gamma^+.
\end{align*}
Note that $\rm \mathbf{E}_m\cdot \mathbf{n}=-[u]/h$. This set of equations have been considered by Miles and Robertson (1932) \cite{miles1932dielectric} for a single sphere. We further assume a thin membrane by setting $\rm h/R_1\rightarrow 0$. 
\begin{figure*}
\subfigure{\includegraphics[width=\linewidth]{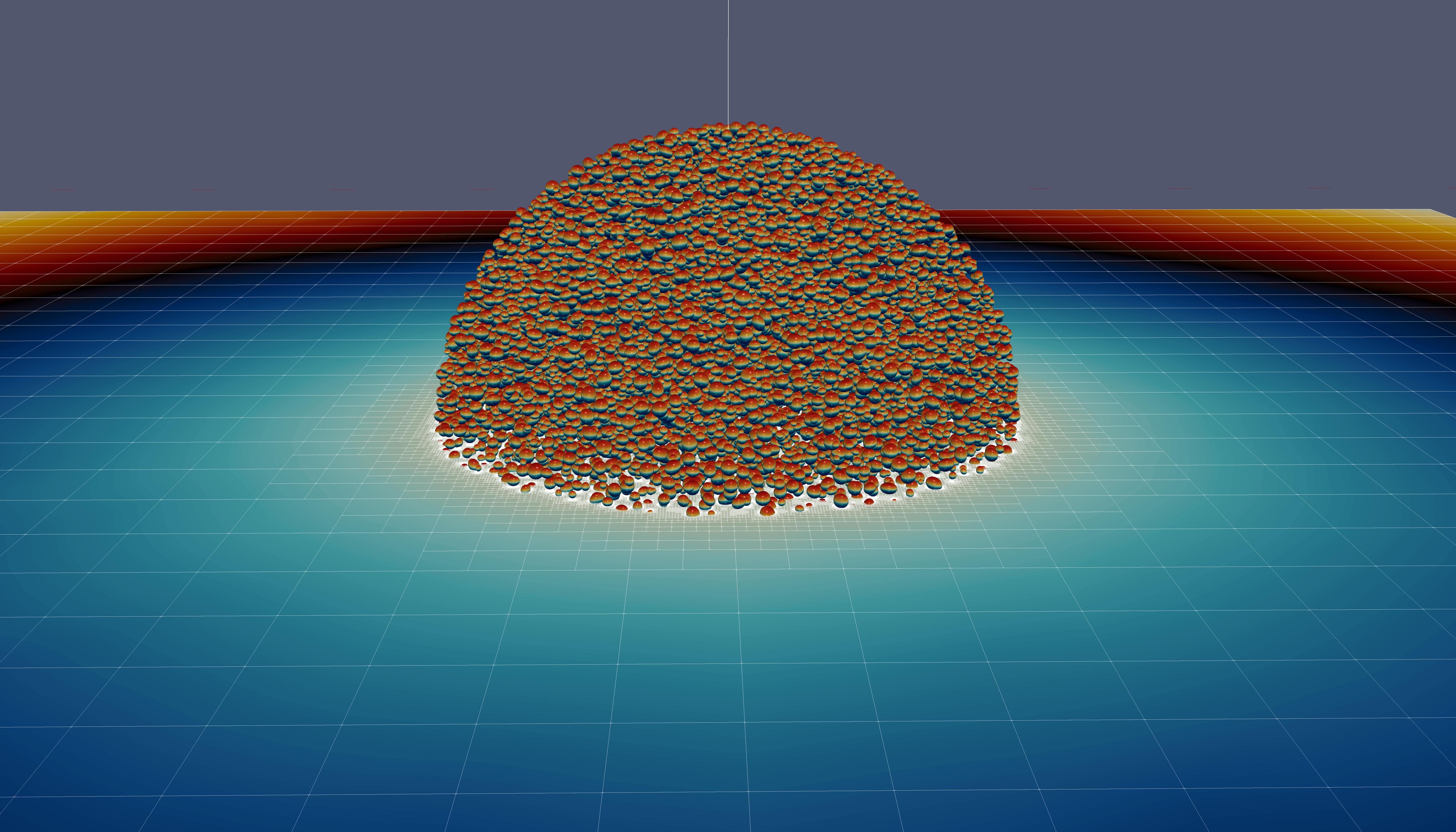} \label{subfig::zoommesh}} \quad \quad %zoom_mesh_TMP
\caption{A snapshot of the $\rm 3D$ spherical tumour composed of $\rm \sim 30,000$ random cells considered in this work. Cells are colored according to their TMP values, with redder colors indicating higher positive TMP while bluer colors indicate lower negative TMP. The Octree data structure is shown on an equatorial slice to emphasize the enhanced resolutions close to cell membranes. }
\label{fig::simuls}
\end{figure*} 
The exact response of the TMP to a step pulse for an isolated sphere with constant membrane conductance has been studied by many authors \cite{kotnik1997sensitivity,gross1988electromobile,Neumann1989,ho1996electroporation,geng2013microfluidic}. Even though these equations are not solved for arbitrary cell geometries, Kotnik \& Miklav\'ci\'c \cite{Kotnik2000} provide analytical solutions for the TMP over oblate and prolate cells in the special case where the cell's axes of symmetry is parallel to the applied field. Qualitatively, they have shown that the maximum TMP increases with the equatorial radius of spheroids, \textit{i.e.} the radius perpendicular to the applied field. The authors considered an insulating membrane to simplify the analysis; however it was shown in experiments that this is not a valid assumption \cite{teissie1993experimental} and one has to consider changes in membrane poration and permeabilization under an applied pulse. This re-structuring of cells membrane under an electric field can be considered, for example, by adopting a nonlinear phenomenological model for the membrane conductance \cite{LEGUEBE201483}:
\begin{equation}
S_m(t,[u]) = S_L +S_1 X_1(t,[u])+S_2 X_2(t,[u]), \label{eq::conductance}
\end{equation}
where  $\rm S_0$, $\rm S_1$ and $\rm S_2$ are the conductance values of the membrane in the resting, porated and permeabilized states, respectively. In this model, the level of poration and permeabilization of the membrane are captured in the functions $\rm X_1$ and $\rm X_2$, which are calculated as a function of the TMP by solving a set of nonlinear ordinary differential equations. Overall, we therefore adopt the following boundary value problem to model electric interactions in charge-free mixtures,
\begin{subequations}
\begin{align}
&\nabla \cdot(\sigma(\mathbf{x}) \nabla u) = 0, & \mathbf{x} \in (\Omega_c \cup \Omega_e) \label{eq::Laplace}
\intertext{with boundary conditions,}
&\left[\sigma(\mathbf{x}) \partial_{n} u\right] = 0 & \mathbf{x} \in  \Gamma, \label{eq::bc1} \\
&C_m \partial_t \left[u\right] + S(t,\left[u\right])\left[u\right] = \sigma(\mathbf{x}) \partial_{n} u & \mathbf{x} \in \Gamma, \label{eq::bc2} \\
&u(t,\mathbf{x}) = g(t,\mathbf{x})   &\mathbf{x} \in \partial\Omega, \label{eq::bc3} 
\intertext{and homogeneous initial condition,}
&u(0,\mathbf{x}) = 0  &\mathbf{x} \in \Omega \label{eq::IC}
\end{align}
\end{subequations}
where we used $ \left[ \boldsymbol{o}\right]=\boldsymbol{o}_e-\boldsymbol{o}_c$ to describe the jump operator across the interface $\rm \Gamma$ in the normal direction, $\rm \sigma_c$, $\rm \sigma_e$ and $\rm \sigma_m$ are the conductivities of the cytoplasm, the extra-cellular medium and the cell membrane, respectively. Note that $\rm \sigma_m\equiv S_m h$ and $\rm \epsilon_0\epsilon_m\equiv C_m h$ are membrane conductivity and permittivity, respectively. 

\subsection{Direct numerical simulations}
Numerical simulations can be used to directly investigate tissue-level properties of cell aggregates emerging from the set of equations \ref{eq::Laplace}--\ref{eq::IC} along with \eqref{eq::conductance}, when considered in a large heterogeneous environment. However, several computational challenges need to be addressed in order to obtain guaranteed accuracy and convergence of the numerical results, while simultaneously considering large enough number of cells. Particularly, imposing jump conditions on numerous irregular sharp interfaces requires efficient numerical discretization methods along with scalable parallel computing algorithms for mesh generation and storage as well as advanced linear system solvers and preconditioners.

In this vein, Guittet \textit{et al.} (2016) \cite{GuittetVoronoi} introduced the Voronoi Interface Method (VIM) to solve Elliptic problems with discontinuities across the interface of irregular domains. Basically, VIM utilizes an interface-fitted Voronoi mesh before applying the dimension-by-dimension Ghost Fluid Method \cite{fedkiw1999non}. Importantly, VIM produces a linear system that is symmetric positive definite with only its right-hand-side affected by the jump conditions. The solution and the solution's gradients are second-order accurate and first-order accurate, respectively, in the $\rm L_\infty$-norm. Later, Guittet \textit{et al.} (2017) applied VIM to the case of cell electroporation \cite{guittet2017voronoi} in a serial computing environment. Recently, Mistani \textit{et al.} (2019) \cite{mistani2019parallel} extended these results to parallel computing environments and considered a large tumour spheroid composed of $\rm \sim 30,000$ ellipsoidal cells. Figure \ref{fig::simuls} illustrates a snapshot of this simulation with the transmembrane potentials depicted over cell membranes. This simulation  leveraged the suite of data structures and routines provided by the Portable, Extensible Toolkit for Scientific Computation (\texttt{PETSc}) \cite{petsc-web-page,petsc-user-ref,petsc-efficient} using the Bi-CGSTAB solver \cite{van1992bi} over the linear system preconditioned by \texttt{hypre} \cite{falgout2002hypre} library. Creation and management of adaptive octree grids was handled by \texttt{p4est} software library \cite{BursteddeWilcoxGhattas11} along with \texttt{voro++} \cite{rycroft2009voro} library for building an adaptive Voronoi tessellation from the underlying Octree grid.

To our knowledge, the direct numerical simulations employed in this work present the current state-of-the-art for large scale simulations of electric interactions at the level of cell aggregates. We believe the simulation results provide precise information about electric response of cell aggregates. In this manuscript we leverage the insights drawn from this direct numerical simulation data to develop and corroborate a theory for the time-dependent evolution of the TMP in cell aggregates under arbitrary applied electric pulse.

\subsection{Multiscale modeling strategy}
Given the microscopic model of the electric interactions \eqref{eq::Laplace}-\eqref{eq::IC}, we aim to infer effective theories at the multicellular level where tens of thousands of cells are present. In our work we focus on the effective properties of the aggregate using the \textit{effective medium theory}, for example see \cite{choy2015effective}, accompanied with a dynamical model for the transient response of the system to an external pulse using the \textit{Fokker-Planck formalism} \cite{fokker1914mittlere,planck1917}. Below, we briefly describe and justify each component of our modeling strategy.

\textbullet\ Effective medium theory: Maxwell (1873) \cite{maxwell1873treatise} was the first to study effective transport properties of a stationary, random and homogeneous suspension of spherical particles dispersed in a background medium of uniform conductivity. Maxwell's assessment of effective conductivity was accurate to order $\mathcal{O}(\phi)$ ($\phi$ is the volume fraction of particles) and relied on his observation that changes in effective conductivity of a suspension of particles was due to the average dipole moment of particles. Exactly a century later, Jeffrey (1973) \cite{jeffrey1973conduction} expanded Maxwell's estimation to order $\mathcal{O}(\phi^2)$ using the general method of Batchelor (1972) \cite{batchelor1972sedimentation,batchelor1972hydrodynamic} by considering pairwise interactions between spherical particles. Batchelor's work was focused on studying the effects of hydrodynamic interactions between particles moving at low velocity through a fluid on the effective viscosity of a suspension of particles. A remarkable result of his work was finding the second order correction term to Einstein's result for the effective viscosity of a suspension of dilute particles by systematically considering pairwise hydrodynamic interactions between freely-moving spheres in a linear flow field.

As part of the work presented here, we apply Batchelor's approach to the problem of conductive shelled spheres instead of the ``homogenization'' method that emerged in the 80's and is more commonly used in similar application domains, \textit{e.g.} in cardiac electrophysiology \cite{franzone2002degenerate,collin2018mathematical}. Our choice is motivated by several reasons: (i) as was pointed out by Hinch (2010) \cite{hinch2010perspective}, homogenization techniques are limited to periodic microstructures, which makes them less applicable for modeling finite size multicellular systems like tumor spheroids, (ii) presenting this problem in Batchelor's formalism allows for application of several existing results such as the influence of the particles' shape and arrangement on the effective conductivity of cell aggregates (even at maximum packing fractions with touching spheres), the consideration of both near- and far-field interactions between particles, as well as the effects due to higher-order multipole moment interactions of particle polarizations on the overall conductivity; we refer the interested reader to Batchelor (1974) \cite{batchelor1974transport}, Bonnecaze \& Brady (1990) \cite{bonnecaze1990method} and the references therein for more details. Lastly, (iii) Batchelor's method is based on ensemble averages of interactions among dispersed particles, that implicitly assumes indistinguishability of particles, which is in line with the Fokker-Planck formalism that we present next.

\textbullet\ Fokker-Planck formalism: Cells in an aggregate are heterogeneous in shapes and electrical properties. Therefore, under an applied electric field, the evolutionary path of a cell's polarization is different from that of other cells, necessitating a probabilistic description of the induced polarizations. To this end, we describe the state of the multicellular system with the probability density of induced dipole moments on membranes and cytoplasms. We then derive a Fokker-Planck equation (FPE) that describes the time evolution of the state-space probability density in response to an electric pulse, as well as the many non-thermal small disturbances that influence the states. Therefore, the Fokker-Planck equation not only provides the stationary state of the system, but also predicts its dynamics far from equilibrium. The Fokker-Planck equation was first used by Fokker (1914) \cite{fokker1914mittlere} and Planck (1917) \cite{planck1917} and have been used to describe numerous systems such as the statistics of laser lights, or the rotations of dipole moments under various potentials. The latter was carried out by Debye who derived a Fokker-Planck equation for the rotational dynamics of polarized molecules and is used to describe the $\gamma$-dispersion discussed in section \ref{Physical_bioelectric_processes}. We refer to Risken (1984) \cite{risken1984fokker} for a standard exposition of this topic, and to \cite{coffey2012langevin} for an overview of applications in sciences and engineering.

The basic idea of our treatment is to modify the boundary conditions on the cell membranes in order to add appropriate disturbances to the system parameters, which in turn provides a Langevin equation for the induced dipole moment. Then, we transform the Langevin equation to its corresponding Fokker-Planck partial differential equation. Finally, we reduce the governing FPE by using a moment-based approach and derive a simple set of ordinary differential equations (ODEs) that captures the evolution of the average and of the variance of induced polarizations. Importantly, the set of ODEs is simple enough that it can be used for real-time predictions and control of transmembrane potentials under arbitrary external electric stimulations and system parameters.

The plan of this manuscript follows: in section \ref{sec::II} we use Green's theorem to decompose cellular polarization into its different components and compute effective conductivity of the medium. In section \ref{sec::III} we use the continuity of flux across cell membranes to develop a Langevin equation for the evolution of membrane polarization, thereafter we perturb the physical parameters in this model and after standard averaging procedure we obtain the corresponding Fokker-Planck equation. We provide an analytic treatment of the governing FPE and leverage a moment based approach to compute the reduced order ODE system for the statistical moments of the induced polarization density. Also, we argue in favor of a fractional order for time derivative in the FPE. Finally, in section \ref{sec::IV} we present numerical results of our model both in the time domain and frequency domain. We conclude our work in section \ref{sec::V}.

\section{Coarse-grained representation}\label{sec::II}

Our strategy is to use a multipole generator scheme to represent the surface current density on individual cell membranes in terms of an equivalent set of dipole moments that reproduce an identical potential in a homogenized medium. 

%Historically, the applications of characterizing electric source terms using multipole generators to physiological research dates back to the original work by Einthoven, Fahr and de Waart (1950) \cite{einthoven1950direction} in the context of electrocardiography, which focused on finding the resultant dipole vector of the human heart. Later, Gabor and Nelson (1954) \cite{gabor1954determination} showed that a resultant dipole moment integrated over the bounding surface of a conducting medium can be used to model a set of current sources and sinks within the volume. However, in their approach the location, magnitude and orientation of the resultant dipoles changed over time making the method impractical. In 1957, Yeh and Martineki introduced an alternative fixed-multipole representation for spherical conductors \cite{yeh1957comparison}, which was then generalized to arbitrary geometries by Geselowitz (1960) \cite{geselowitz1960multipole}. 

\subsection{Cells as arrays of layer potentials}
We denote by $\rm u(\mathbf{r})$, the electric potential at any point $\rm \mathbf{r}$ within the aggregate; $\rm u(\mathbf{r})$ satisfies Laplace's equation. We treat this problem by viewing a membrane as a dipole layer (\textit{cf.} see chapter 1 of \cite{jackson1999classical} for more details) that is formed by two infinitesimally close surfaces with opposite charge densities, as depicted in figure \ref{fig::bilayer}. In this work we only consider a passive environment, \textit{i.e.} that is free from current source or sink terms on cell membranes, and denote the passive current by $\rm \mathbf{J}_m$. In this case, we can relate the normal current density passing through the membrane by $\rm \mathbf{J}_m^\pm\cdot\mathbf{n}$ (where $\pm$ refers to the external and internal side of the membrane) to the potential jump across membrane by writing
\begin{align}
-\mathbf{J}_m^\pm\cdot \mathbf{n}=\sigma_c\partial_{\mathbf{n}} u_c=\sigma_e\partial_{\mathbf{n}} u_e, \label{eq::assume}
\end{align}

Similar to Geselowitz \cite{geselowitz1967bioelectric}, we use Green's theorem to treat this problem in terms of source current densities and applied electric fields surrounding the domain. Consider two well-behaved functions $\psi$ and $\phi$ defined inside and outside of cells and define the vector field $\mathbf{F}=\sigma\psi\nabla \phi$ such that it is a continuous function of position in enclosed volumes between boundaries. Substituting $\mathbf{F}$ in Gauss's theorem and wrapping thin membranes with discontinuity in between two close surfaces, one obtains Green's theorem for non-homogeneous mixtures with discontinuity across internal boundaries (see page 53 of \cite{SmytheWilliamRalph1967Sade}),
\begin{align*}
&\sum_{j=1}^q \int_{A_j}[\sigma_c(\psi_c \nabla \phi_c - \phi_c\nabla\psi_c) -\sigma_e(\psi_e \nabla \phi_e - \phi_e\nabla\psi_e)]\cdot d\mathbf{A}_j\\
&+\sum_{j=1}^p \int_{A_j}\sigma ( \psi \nabla \phi -\phi\nabla \psi )\cdot d\mathbf{A}_j\\
&=\int_{V} [\psi \nabla\cdot( \sigma \nabla \phi) - \phi \nabla\cdot (\sigma \nabla \psi)]dv
\end{align*}
where $q$ is the number of surfaces across which $\sigma$ is discontinuous, $p$ is the number of additional boundaries including macroscopic boundaries where electrodes are located, and $V$ is the volume enclosed between $\Gamma$ and all inner surfaces $A_j$ excluding surfaces of discontinuity. Here we adopt a convention that $d\mathbf{A}_j$ is the measure of an area element of the surface $A_j$ that always points into the extra-cellular matrix.

Geselowitz, in his case \textit{I}, considers $\psi=1/r$ and $\phi= u$ with $r$ being the distance between surface or volume elements to any arbitrary point in the domain. Then it is straightforward to show that for any observation point $\mathbf{x}$ in the aggregate, we have
\begin{align*}
4\pi u(\mathbf{x})&=\sum_{m}\int_{A_m} (\sigma_e u_e - \sigma_c u_c) \nabla' (\frac{1}{r})\cdot d\mathbf{A} \nonumber \\
& + \int_{\Gamma}\bigg(\frac{\mathbf{J}_b}{\sigma_e r}+ u_b\nabla' \frac{1}{r} \bigg)\cdot d\mathbf{A}, 
\end{align*}
where $\Gamma$ is the surface of the electrodes, $u_b$ is the electric potential applied at the electrodes, $\mathbf{J}_b=-\sigma_e\nabla u$, and $A_m$ is the surface of membranes that points into the extra-cellular matrix. Moreover, as in case \textit{II} of Geselowitz, we could let $\psi=1/r$ and $\sigma\phi= u$ to show that for infinitesimally thin membranes the electric potential is given by:
\begin{align}
4\pi u(\mathbf{x})&= 
\sum_{m}\int_{A_m} \bigg[\bigg( \frac{1}{\sigma_e} - \frac{1}{\sigma_c}\bigg)\frac{\mathbf{J}_m^\pm}{r} + [u]\nabla' (\frac{1}{r})\bigg]\cdot d\mathbf{A} \nonumber \\
& + \int_{\Gamma}\bigg(\frac{\mathbf{J}_b}{\sigma_e r}+ u_b\nabla' \frac{1}{r} \bigg)\cdot d\mathbf{A}, \label{eq::genformalism}
\end{align}
Hence, using the divergence theorem one can obtain the alternative formulation:
\begin{align}
4\pi u(\mathbf{x})&=\sum_{m}\int_{A_m} [u]\nabla' (\frac{1}{r})\cdot d\mathbf{A}_m \nonumber  \\
& + \sum_c \int_{V_c} \bigg( \frac{1}{\sigma_e} - \frac{1}{\sigma_c}\bigg) \mathbf{J}\cdot\nabla'  (\frac{1}{r}) dV \nonumber \\
& + \int_{\Gamma}\bigg(\frac{\mathbf{J}_b}{\sigma_e r}+ u_b\nabla' \frac{1}{r} \bigg)\cdot d\mathbf{A}, \label{eq::genformalism2}
\end{align}
where $\mathbf{r}=|\mathbf{x}-\mathbf{x}'|$, $\nabla'=\partial/\partial \mathbf{x}'$ and the summation is over the membranes of the enclosed cells within $\Gamma$. The first term on the right-hand side describes the influence of the polarized membranes, the second term captures the contribution from the cytoplasms, and the last term represents the influence of electrodes. We also emphasize that if the electrodes are not in direct contact with the extra-cellular matrix (e.g. there is a gap with low conductivity between $\Gamma$ and the outer surface of the aggregate), one has to also include the extra contribution from the outer surface ($\mathbf{A}_o$),
\begin{align*}
u_o(\mathbf{x})= \frac{1}{4\pi}\int_{A_o}\frac{2\mathbf{E}_o}{r}\cdot d\mathbf{A}.
\end{align*}
In addition to the contribution from the electrodes, equation \eqref{eq::genformalism} decomposes the electric potential at any point in the volume as a superposition of a monopole/single layer and a dipole layer on each membrane. Remarkably, the membrane integral over transmembrane jump is analogous to the contribution from a surface current dipole density:
\begin{align*}
u_p(\mathbf{x})&=\sum_m\frac{1}{4\pi \sigma_e}\int_{\mathbf{A}_m}D(\mathbf{x}') \nabla' \bigg( \frac{1}{r}\bigg) \cdot d\mathbf{A}_m
\end{align*} 
where the current dipole surface density (\textit{i.e.} current times distance between sources) is defined by:
\begin{align*}
D(\mathbf{x})&= \sigma_e [u]
\end{align*}
Then, a point dipole on the membrane is expressed as $\delta \boldsymbol{P}=\sigma_e  [u] d\mathbf{A}_m$ and one may model the induced transmembrane potential as a resultant current dipole on each cell.Thus, we approximate each membrane with surface $A_i$ by a resultant dipole of strength
\begin{align}
\mathbf{P}_i &= \int_{A_i} \sigma_e  [u] d\mathbf{A}, \label{eq::dipolemoment}
\end{align}
and we call $\mathbf{P}$ the \textit{dipolar polarization}. 

Furthermore, we observe that the membrane integral over the transmembrane current density resembles the contribution from a \textit{monopole current layer},
\begin{align*}
u_s(\mathbf{x})&=\sum_m\frac{1}{4\pi }\int_{\mathbf{A}_m} \bigg( \frac{1}{\sigma_e} - \frac{1}{\sigma_c} \bigg)\frac{\mathbf{J}^\pm_m}{r} \cdot d\mathbf{A}_m \nonumber \\
&=\sum_m\frac{1}{4\pi \sigma_e}\int_{\mathbf{A}_m} (\sigma_c -\sigma_e )\frac{\mathbf{E}_c}{r} \cdot d\mathbf{A}_m
\end{align*}
where $\mathbf{E}_c$ refers to the electric field at the inner surface of the membrane. Here, we identify an induced polarization density of $(\sigma_c-\sigma_e)\mathbf{E}_c\cdot\mathbf{n}/(4\pi\sigma_e)$ over cell membranes. Alternatively, this term in its volume integral form reads
\begin{align*}
u_s(\mathbf{x})&=\sum_c\frac{1}{4\pi \sigma_e}\int_{V_c} (\sigma_c - \sigma_e)\mathbf{E}\cdot\nabla' \big(\frac{1}{r}\big)dV,
\end{align*}
which can be interpreted as an `extra flux density', denoted by $\boldsymbol{\tau}$ and is zero everywhere in the extra-cellular matrix while in the cells is given by
\begin{align*}
\boldsymbol{\tau}(\mathbf{x})&=(\sigma_c - \sigma_e)\mathbf{E}\\
&=\mathbf{J}(\mathbf{x}) + \sigma_e \nabla u(\mathbf{x}).
\end{align*}
Hence
\begin{align*}
u_s(\mathbf{x})&=\sum_c\frac{1}{4\pi \sigma_e}\int_{V_c} \boldsymbol{\tau}(\mathbf{x}')\cdot\nabla' \big(\frac{1}{r}\big)dV, 
\end{align*}
which again resembles the electric potential of a volume dipole density $\boldsymbol{\tau}$. Therefore, we define the \textit{instantaneous polarization} ($\mathbf{S}$) to model the polarization of cell cytoplasms:
\begin{align*}
\mathbf{S}_i=\int_{V_i}\boldsymbol{\tau}dV,
\end{align*}
or, equivalently, in terms of the potential in the cytoplasmic side of the membrane as
\begin{align*}
\mathbf{S}_i&=(\sigma_e - \sigma_c)\int_{A_i} u_c ~d\mathbf{A}.
\end{align*}
Then the net polarization can be defined by $\mathbf{M}_i$ 
\begin{align*}
\mathbf{M}_i &=\mathbf{P}_i+\mathbf{S}_i=\int_{A_i}(\sigma_e u_e - \sigma_c u_c)~d\mathbf{A}
\end{align*}
We also note that this result could be directly inferred using Green's theorem by letting $\phi=u$ and $\psi=1/r$, \textit{e.g.} see equation 29 of Geselowitz 1967. Furthermore one can relate the values of $\mathbf{P}$ and $\mathbf{S}$ through the equation:
\begin{align}
\frac{\mathbf{P}}{\sigma_e}=\int_{A}u_ed\mathbf{A}+\frac{\mathbf{S}}{\sigma_c -\sigma_e} \label{eq::relationSP}
\end{align}
We observe that in general the dipolar and instantaneous polarizations may have different orientations depending on the symmetries of the exterior potential. To assess this relation, we use Gauss' theorem for a closed surface $\Gamma$ enclosing $N$ internal closed surfaces (\textit{cf.} see chapter III of Smythe, note the minus sign is due to our convention that the normal direction points into the extra-cellular matrix),
\begin{align}
\sum_{j=1}^{N} \int_{A_j}u~ d\mathbf{A} + \int_{\Gamma} u ~d\mathbf{A} = -\int_{V_e} \nabla u ~ dV, \label{eq::Gauss}
\end{align}
where $V_e$ is the volume of the extra-cellular matrix excluding cells. Therefore, for the cell aggregate
\begin{align}
\frac{n<\mathbf{P}>}{\sigma_e}=\frac{\sum_{j=1}^N \int_{A_j}u ~ d\mathbf{A}}{V} + \frac{n<\mathbf{S}>}{\sigma_c -\sigma_e}, \label{eq::rell}
\end{align}
where,
\begin{align*}
&n<\mathbf{P}>=\frac{\sum_{j=1}^N\mathbf{P}_j}{V}\quad \textrm{and} \quad n<\mathbf{S}>=\frac{\sum_{j=1}^N\mathbf{S}_j}{V},
\end{align*}
with $n$ the number density of cells in the mixture. Note that the volume fraction $\phi$ is related to the number density $n$ via $\phi=n V_c$.  Using Gauss law \eqref{eq::Gauss} with equation \ref{eq::rell} we obtain
\begin{align*}
\frac{n<\mathbf{P}>}{\sigma_e}&=\frac{n<\mathbf{S}>}{\sigma_c -\sigma_e} - \frac{1}{V}\int_{V_e}\nabla u~ dV - \frac{1}{V}\int_{\Gamma}u d\mathbf{A}. 
\end{align*}
Furthermore, we define the applied electric pulse on the boundary, $\mathbf{E}_{ext}$, as:
\begin{align*}
\mathbf{E}_{ext} \equiv \frac{1}{V}\int_{\Gamma}u d\mathbf{A}.
\end{align*}
We also recall that $\mathbf{S}$ is related to the volume averaged electric field in cell cytoplasms by:
\begin{align*}
\phi \bar{\mathbf{E}}_c &\equiv -\frac{1}{V}\sum_{j}^{cells}\int_{V_j}\nabla u~dV = \frac{n<\mathbf{S}>}{\sigma_c - \sigma_e},
\end{align*}
while the volume averaged external field is simply
\begin{align*}
(1-\phi) \bar{\mathbf{E}}_e &\equiv - \frac{1}{V}\int_{V_e}\nabla u~ dV \nonumber \\
&=\mathbf{E}_{ext} + \frac{n<\mathbf{P}>}{\sigma_e} + \frac{n<\mathbf{S}>}{\sigma_e - \sigma_c} 
\end{align*}
and the volume average electric field inside the membranes is related to the dipolar polarization by
\begin{align*}
\frac{1}{V}\sum_{j}^{cells}\int_{V'_j}\nabla u dV =\frac{1}{V} \sum_{j}^{cells} \int_{\Gamma_j} [u]d\mathbf{A} =\frac{n<\mathbf{P}>}{\sigma_e}.
\end{align*}
Because the volume $V$ is partitioned into three parts $V = V_e \cup V_c \cup V'_c$, where $V'_c$ is the volume occupied by cell membranes, we conclude that
\begin{align*}
-<\nabla u >&=\mathbf{E}_{ext}  \\
&= \phi\bar{ \mathbf{E}}_c +  (1-\phi) \bar{ \mathbf{E}}_e -  \frac{n<\mathbf{P}>}{\sigma_e}.
\end{align*}
In particular, we note that $\phi V=\sum_j V_j$, and we define the dipolar polarization per cell volume as $\mathbf{p}_j=\frac{\mathbf{P}_j}{V_j}$ to obtain the effective dipole moment per cell volume,
\begin{align*}
\bar{\mathbf{p}}=\frac{\sum_j V_j \mathbf{p}_j}{\sum_j V_j}, 
\end{align*}
and we can write
\begin{align*}
n<\mathbf{P}> =  \phi \bar{\mathbf{p}}.  
\end{align*}

\subsection{Frequency domain model for cell dipoles}
We consider the analytical solution of a spherical cell of radius $R_1$ centered within a spherical domain of radius $R_2$ under a Dirichlet potential $E(t)R_2\cos\theta$ at the outer boundary (a \textit{local} electric field $\mathbf{E}=-E\mathbf{k}$ is considered at the surface of a sphere of radius $r=R_2$ from center of the cell). In this case the membrane voltage satisfies
\begin{align*}
&C_m\frac{\partial [u]}{\partial t}+(S_L -B)[u]=A E(t) R_2\cos\theta, 
\end{align*}
where
\begin{align*}
&A=3\sigma_c\sigma_e R_2^2K \quad \textrm{and} \quad B=-\sigma_c\sigma_e(R_1^2 + 2\frac{R_2^3}{R_1})K, 
\end{align*}
with $K^{-1}=R_1^3(\sigma_e - \sigma_c) + R_2^3(2\sigma_e + \sigma_c)$ so the coefficients are:
\begin{align*}
&A=\frac{3\sigma_c \sigma_e/R_2}{ 2\sigma_e + \sigma_c + \phi (\sigma_e - \sigma_c)}, \\
\textrm{and} \\
&B=-\frac{(2+\phi)\sigma_c \sigma_e/R_1}{ 2\sigma_e + \sigma_c + \phi (\sigma_e - \sigma_c)},
\end{align*}
where the volume fraction of cells is given by $\phi = R_1^3/R_2^3$. Furthermore, we define three independent parameters that characterize the solution:
\begin{align*}
&\tilde{\sigma}=  2\sigma_e + \sigma_c + \phi (\sigma_e - \sigma_c),\\
&\eta=1 + \frac{S_L R_1 \tilde{\sigma}}{(2+\phi)\sigma_c\sigma_e},\\
&\tau=\frac{ \tilde{\sigma} R_1 C_m}{(2+\phi)\sigma_c\sigma_e}.
\end{align*}
In the frequency domain the solutions are given by,
\begin{align*}
&[\tilde{u}]=\frac{3R_1 }{2+\phi}\cdot \frac{1}{\eta + j\omega\tau}\cdot \tilde{E}(\omega) \cdot \cos\theta \\
&\tilde{u}_c = \tilde{\alpha}_c  \tilde{E}(\omega)\cdot r \cdot \cos\theta\\
&\tilde{u}_e = (\tilde{\alpha}_e r + \frac{\tilde{\beta}_e}{r^2})\cdot \tilde{E}(\omega) \cdot \cos\theta
\end{align*}
with
\begin{align*}
&\tilde{\alpha}_c = \frac{3\sigma_e}{\tilde{\sigma}}\cdot \bigg( \frac{\eta - 1 + j\omega\tau}{\eta + j\omega\tau}\bigg)   \\
&\tilde{\alpha}_e =\bigg( \frac{\sigma_c + 2\sigma_e}{\tilde{\sigma}} - \frac{3\sigma_c }{\tilde{\sigma}} \cdot \frac{\phi}{2+\phi} \cdot \frac{1}{\eta + j\omega\tau}\bigg) \\
&\tilde{\beta}_e =\bigg(\frac{\sigma_e - \sigma_c}{\tilde{\sigma}} + \frac{3\sigma_c}{\tilde{\sigma}}\cdot \frac{1}{2+\phi}\cdot\frac{1}{\eta+j\omega\tau}  \bigg)\cdot R_1^3
\end{align*}
The external and internal electric fields are computed given the electric potential ($\mathbf{r}=r\hat{r}$)
\begin{align}
\tilde{\mathbf{E}}_e (\omega) &= \tilde{\alpha}_e   \tilde{ \mathbf{E}}(\omega) - \frac{3 (\tilde{\beta}_e \tilde{\mathbf{E}}(\omega)\cdot \hat{r})  \hat{r} - \tilde{\beta}_e  \tilde{ \mathbf{E}}(\omega) }{r^3}, \label{eq::Ee}\\
\textrm{and} \nonumber \\
\tilde{\mathbf{E}}_c (\omega) &= \tilde{\alpha}_c \tilde{\mathbf{E}}(\omega),
\end{align}
which corresponds to a uniform external field superimposed by the electric field of a net dipole moment 
\begin{align}
\tilde{\mathbf{M}}(\omega)=-4\pi\sigma_e \tilde{\beta}_e \tilde{ \mathbf{E}}(\omega). \label{eq::MeDirect}
\end{align}
It can be easily verified that indeed $u_e(R_2)=ER_2\cos\theta$ as expected. We emphasize that we only impose the tangential component of electric field at $r=R_2$ and not its radial component, as evident by equation \eqref{eq::Ee}.

As we discussed before, we represent this solution by defining dipole moments over membranes and cytoplasms. It is straightforward to compute the dipole moments from the basic definitions of the previous section, which lead to
\begin{align}
\frac{\tilde{\mathbf{P}}(\omega)}{V_c}&=-\frac{3\sigma_e}{2+\phi} \cdot \frac{1}{\eta + j\omega\tau}\cdot \tilde{\mathbf{E}}(\omega) \label{eq::explicitP}\\
\textrm{and} \nonumber \\
\frac{\tilde{\mathbf{S}}(\omega)}{V_c}&=-\frac{3\sigma_e (\sigma_e - \sigma_c)}{\tilde{\sigma}}\cdot \frac{\eta - 1 + j\omega\tau}{\eta + j\omega\tau}\cdot \tilde{\mathbf{E}}(\omega), \label{eq::explicitS}
\end{align}
where $V_c$ is the volume of a cell. One can verify that indeed we have $\tilde{\mathbf{P}}+\tilde{\mathbf{S}} \equiv \tilde{\mathbf{M}}$ from equation \eqref{eq::MeDirect}. Moreover, it is straightforward to verify that $\tilde{\mathbf{P}}$ and $\tilde{\mathbf{S}}$ are related through equation \eqref{eq::relationSP}. We note the minus sign in dipole moments stems from our mathematical definition for jump in solution across membrane, that is the value of solution in the exterior minus that of the cytoplasm, which is opposite the usual convention in biology. Therefore, in making the connection with other works, care must be taken in using consistent signs at this step. Note also that the influence of the membrane conductivity is captured in the parameter $\eta$. For example, for an insulating membrane where $\sigma_m\ll \sigma_c$, we have $\eta\approx 1$ and therefore the cell cytoplasm is effectively shielded from polarization in agreement with experiments, \textit{i.e.} $\mathbf{S}=0$. 
 
\begin{figure*}
\subfigure{\includegraphics[width=0.45\linewidth]{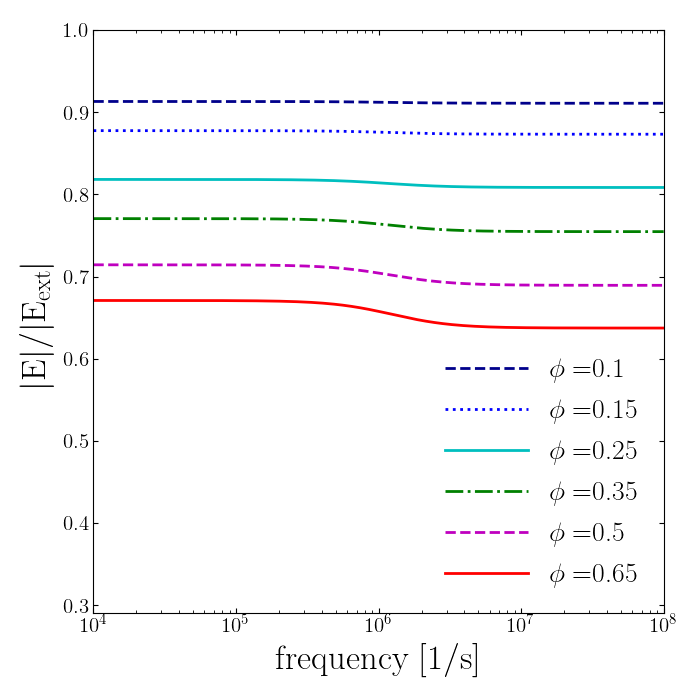}}
\subfigure{\includegraphics[width=0.45\linewidth]{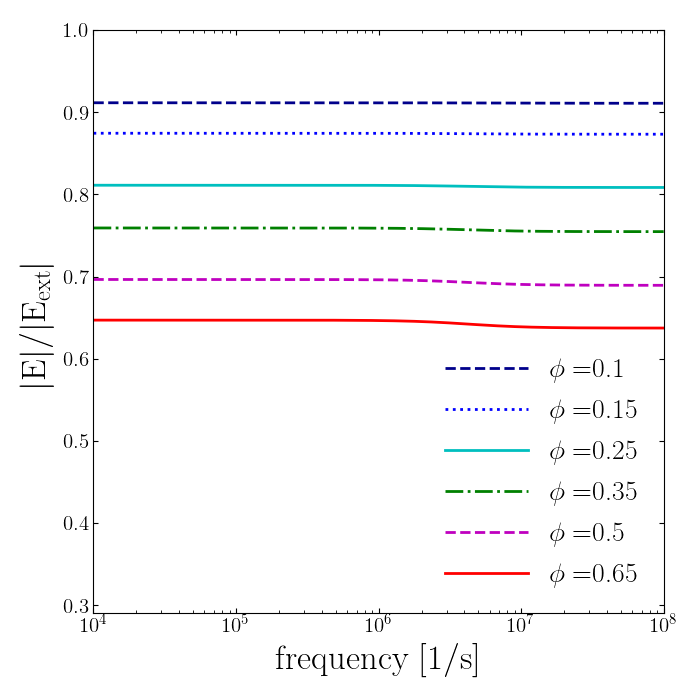}}
\caption{The magnitude of the local electric field at different frequencies and volume fractions for $(\sigma_e,\sigma_c)=(1.3, 0.6) \ [S/m]$ and $S_L=1.9\ [S/m^2]$ (left) and $S_L=1.9\times 10^5\ [S/m^2]$ (right).}
\label{fig:Eextracellular}
\end{figure*}

Equations \eqref{eq::explicitP}--\eqref{eq::explicitS}  enable the definition of the cellular polarizability coefficients $\mathbf{P}=\sigma_e\alpha_p \mathbf{E}$, $\mathbf{S}=\sigma_e\alpha_s \mathbf{E}$, and $\mathbf{M}=\sigma_e\alpha \mathbf{E}$ with
\begin{align*}
\alpha_{p} &= -\frac{3}{2+\phi} \cdot \frac{1}{\eta + j\omega\tau}, \\
\alpha_{s} &= -\frac{3 (\sigma_e - \sigma_c)}{\tilde{\sigma}}\cdot \frac{\eta - 1 + j\omega\tau}{\eta + j\omega\tau}, \\
\textrm{and} \nonumber \\
\alpha  &= \alpha_p + \alpha_s. 
\end{align*}
Also note that $\tilde{\alpha}_e$ is related to the electric polarizability $\alpha_p$ via
\begin{align*}
\tilde{\alpha}_e&=\frac{ 2\sigma_e + (1+\phi\alpha_p)\sigma_c }{\tilde{\sigma}}. 
\end{align*}
The average polarization of the whole aggregate is given by:
\begin{align}
n<\tilde{\mathbf{P}}>&= \phi \cdot \alpha_p \cdot \sigma_e\tilde{\mathbf{E}} \label{eq::aggregateP}\\
n<\tilde{\mathbf{S}}>&=  \phi \cdot \alpha_s \cdot \sigma_e\tilde{\mathbf{E}}\label{eq::aggregateS}
\end{align}
Importantly, $\tilde{\mathbf{E}}$ can be related to the applied pulse $\tilde{\mathbf{E}}_{ext}$ by averaging equation \eqref{eq::Ee} within a spherical shell volume between the membrane and $R_2$. By integration, the dipolar contribution is zero, and we establish a relationship with the average external electric field, $\bar{\mathbf{E}}_{e}$, $\tilde{\alpha}_e \tilde{\mathbf{E}} = (1-\phi) \bar{\mathbf{E}}_{e}$. We already showed that
\begin{align*}
(1-\phi) \bar{\mathbf{E}}_{e}&=\tilde{\mathbf{E}}_{ext} + \frac{n<\tilde{\mathbf{P}}>}{\sigma_e} + \frac{n<\tilde{\mathbf{S}}>}{\sigma_e-\sigma_c},
\end{align*}
therefore the local electric field is given by
\begin{align}
\tilde{\mathbf{E}} &= \kappa^{-1}\tilde{\mathbf{E}}_{ext}, \label{eq::E_by_Eext}\\
\textrm{with} \nonumber \\
\kappa &= \alpha_e -\phi\alpha_p - \phi \frac{\sigma_e\alpha_s}{\sigma_e - \sigma_c},
\end{align}
which simplifies to
\begin{align}
\kappa =  \frac{(2+3\phi)\sigma_e + \sigma_c }{\tilde{\sigma}} - \frac{3\phi^2\sigma_c}{(2+\phi)\tilde{\sigma}(\eta + j\omega\tau)}.\label{eq::kappa}
\end{align}
For example figure \ref{fig:Eextracellular} illustrates the magnitude of $\mathbf{E}$ for two different membrane conductivities.

At the aggregate level, we can define the membrane susceptibility, $n<\mathbf{P}>=\sigma_e\chi_{p} \mathbf{E}_{ext}$, the cytoplasm susceptibility, $n<\mathbf{S}>=\sigma_e\chi_{s}\mathbf{E}_{ext}$, as well as the overall cell susceptibility, $n<\mathbf{M}>=\sigma_e \chi \mathbf{E}_{ext}$, that relate the applied electric pulse to the induced polarization densities. 
Equation \eqref{eq::E_by_Eext} allows us to relate the applied pulse to the induced polarizations, therefore we obtain the susceptibility coefficients
\begin{align*}
&\chi_{p} = \frac{\phi \alpha_p}{\kappa} \quad \textrm{and} \quad \chi_{s} = \frac{\phi \alpha_s}{\kappa},
\end{align*}
which also imply that
\begin{align*}
&\frac{(1-\phi)\bar{E}_{e}}{E_{ext}}=\frac{\alpha_e}{\kappa}  \quad \textrm{and} \quad \frac{\phi \bar{E}_c}{E_{ext}}=\frac{\sigma_e\chi_s}{\sigma_c - \sigma_e}.
\end{align*}

The time-domain version of $\mathbf{E}$ can be found by expressing the complex factor $\eta + j\omega\tau$ in equation \eqref{eq::kappa} in terms of $<\mathbf{P}>$. By noting that
\begin{align*}
\eta + j\omega\tau = \frac{-3\phi \sigma_e \tilde{E}}{(2+\phi)n<P>}, 
\end{align*}
we obtain that
\begin{align*}
\kappa = \frac{(2+3\phi)\sigma_e +\sigma_c}{\tilde{\sigma}} + \phi\frac{\sigma_c}{\tilde{\sigma}}\frac{n<\tilde{P}>}{\sigma_e \tilde{E}},
\end{align*}
which, substituted into equation \eqref{eq::E_by_Eext} leads to
\begin{align*}
\mathbf{E}(t)=\frac{\tilde{\sigma}\mathbf{E}_{ext}(t) - \nu \phi n <\mathbf{P}(t)>}{(2+3\phi)\sigma_e + \sigma_c}
\end{align*}
where $\nu = \sigma_c/\sigma_e$. For the purpose of developing time-domain equations it is also useful to write the cytoplasm polarization in terms of the membrane polarization:
\begin{align*}
n<\mathbf{S}>=-3\phi \frac{\sigma_e - \sigma_c}{\tilde{\sigma}} \bigg( \sigma_e \mathbf{E} + \frac{2+\phi}{3\phi}n<\mathbf{P}>\bigg).
\end{align*}

\subsection{Effective conductivity}
Maxwell (1873) \cite{maxwell1873treatise} first considered the problem of calculating effective conductivity coefficients for dilute spherical inclusions. A century later, Jeffrey (1973) \cite{jeffrey1973conduction} included pairwise interactions into Maxwell's theory, for increasing the validity of the estimated effective conductivity for higher concentrations. Chiew and Glandt (1983) \cite{chiew1983effect} considered a more realistic pair-correlation function to improve on the accuracy of Jeffrey's result. In parallel, Hasselman and Johnson 1987 \cite{hasselman1987effective} included the effect of interfacial resistance to Maxwell's theory, which was subsequently integrated with Jeffrey's theory by Chiew and Glandt \cite{chiew1987effective}. In this section, we derive the effective conductivity based on the work of Batchelor for transport phenomena in two-phase media composed of an statistically homogeneous suspension of particles with random configurations (see \textit{e.g.} \cite{batchelor1974transport,o1979method,bonnecaze1990method}). We emphasize that for improved accuracy one has to modify this approach to include divergences of all higher order multipole moments; we refer the interested reader to \cite{dahler1963theory,russakoff1970derivation}.

First, we define the average flux in a volume large enough to include many cells,
\begin{align*}
&<\mathbf{J}>=\frac{1}{V}\int_V \mathbf{J} ~ dV,
\end{align*}
and we seek a linear relationship between the average flux and the potential gradient
\begin{align}
<\mathbf{J}>=-\bar{\sigma}<\nabla u>, \label{eq::HM5}
\end{align}
where the proportionality coefficient defines the \textit{effective conductivity}. We decompose $<\mathbf{J}>$ into three different regions,
\begin{align*}
<\mathbf{J}>&=\frac{1}{V}\int_{V-\sum_i V_i}\mathbf{J}dV + \frac{1}{V}\sum_i \int_{V_i\cup V'_i} \mathbf{J}dV\\
&=\frac{1}{V}\int_{V}-\sigma_e \nabla u ~ dV + \frac{1}{V}\sum_i \int_{V_i\cup V'_i} \boldsymbol{\tau} dV
\end{align*}
where $\sum_i V_i$ is the volume occupied by cells. Last expression is obtained by replacing $\mathbf{J}_k=-\sigma_k\nabla u \equiv -\sigma_e\nabla u + \boldsymbol{\tau}_k$ such that $\mathbf{\tau}_k=(\sigma_e-\sigma_k) \nabla u$. In the membrane we approximate $\nabla u=[u]\mathbf{n}/h$, therefore,
\begin{align}
<\mathbf{J}>&=-\sigma_e<\nabla u> + n <\mathbf{S}> + \big(1 - \frac{\sigma_m}{\sigma_e}\big)n<\mathbf{P}> \label{eq::mainCurrent}
\end{align}
which in terms of the dipolar polarization and the external electric field is given by
\begin{align*}
<&\mathbf{J}>=\sigma_e \mathbf{E}_{ext} \bigg(\frac{2+\nu + 3\phi\nu}{2+ \nu + 3\phi}\bigg) + \nonumber \\
& n<\mathbf{P}>\bigg( 1 - \frac{\sigma_m}{\sigma_e}- \frac{\sigma_e - \sigma_c}{\tilde{\sigma}}[ 2 + \phi - \frac{3\nu \phi^2}{2+ 3\phi + \nu} ] \bigg).
\end{align*}
Moreover, using the definition of the effective conductivity \eqref{eq::HM5} and the fact that $-<\nabla u>=\mathbf{E}_{ext}$, we deduce that the parallel ($\parallel$) and the transverse ($\perp$) components of the effective conductivity are given by
\begin{align*}
&\frac{\bar{\sigma}_\parallel}{\sigma_e} = \frac{ <J_\parallel> }{\sigma_e E_{ext} } \quad \textrm{and} \quad \frac{\bar{\sigma}_\perp}{\sigma_e} = \frac{ <J_\perp> }{\sigma_e E_{ext} }.
\end{align*}
Particularly, in the frequency domain, the parallel component satisfies
\begin{align}
\frac{\bar{\sigma}}{\sigma_e} = 1 + \chi_s +  (1 - \frac{\sigma_m}{\sigma_e}) \chi_p. \label{eq:homogenization}
\end{align}
For infinitely conductive membranes, where $\eta\rightarrow\infty$, equation \eqref{eq:homogenization} can be simplified by noting that 
\begin{align}
\frac{\sigma_m}{\sigma_e}= \frac{h}{R_1} \cdot \frac{(2+\phi)\sigma_c}{\tilde{\sigma}}\cdot  (\eta - 1)
\end{align}
and that the membrane polarization vanishes. In this case, equation \eqref{eq:homogenization} reduces to Maxwell's equation for the effective conductivity of a dilute suspension \cite{maxwell1881treatise}:
\begin{align}
\frac{\bar{\sigma}}{\sigma_e}\rightarrow 1 - 3\phi \frac{1-\nu + \nu h/R_1}{2+\nu + 3\phi}\label{eq::Maxwell}
\end{align}
and $h/R_1\rightarrow 0$. Note that Maxwell's result is only a $\mathcal{O}(\phi)$ estimate of the effective conductivity, and that, to its limit of validity, equation \eqref{eq::Maxwell} coincides with Maxwell's approximation (see \textit{e.g.} \cite{jeffrey1973conduction}).

\begin{figure*}
\subfigure[]{\includegraphics[height=0.4\linewidth]{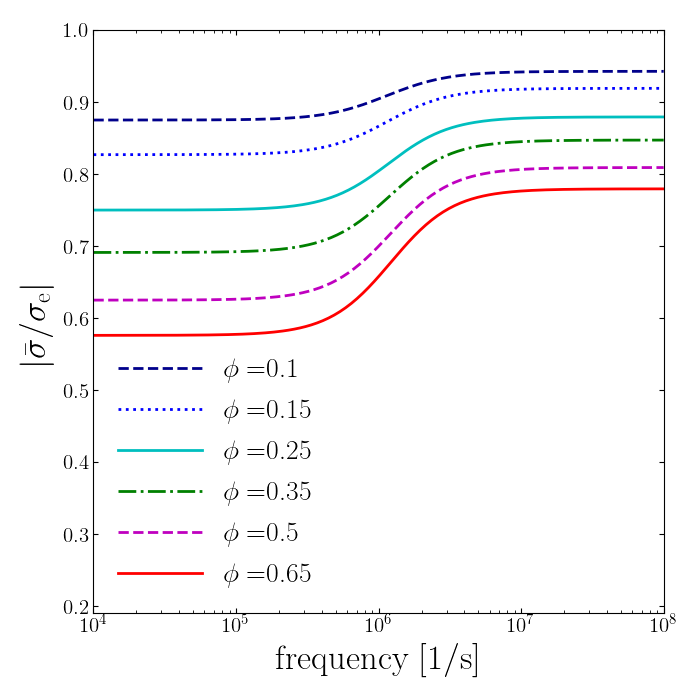}\label{fig:homogenization1}}
\subfigure[]{\includegraphics[height=0.4\linewidth]{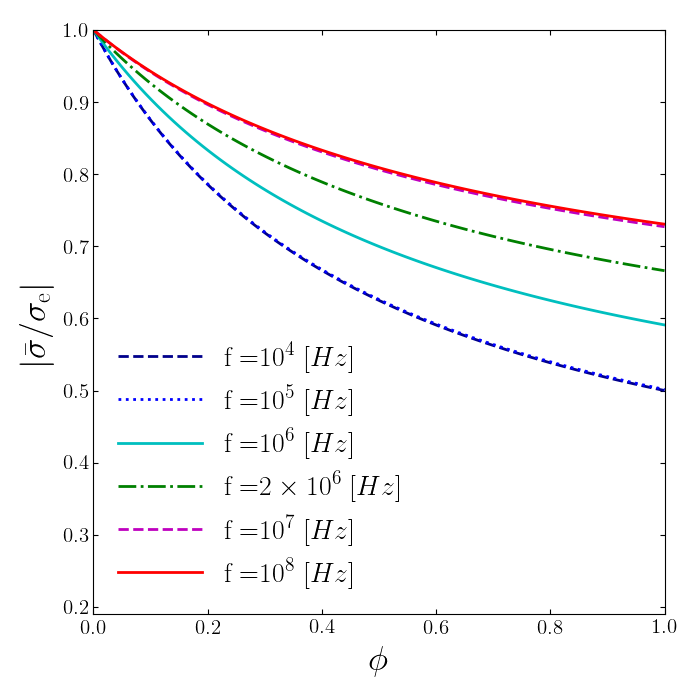}\label{fig:homogenization12}}
\subfigure[]{\includegraphics[height=0.4\linewidth]{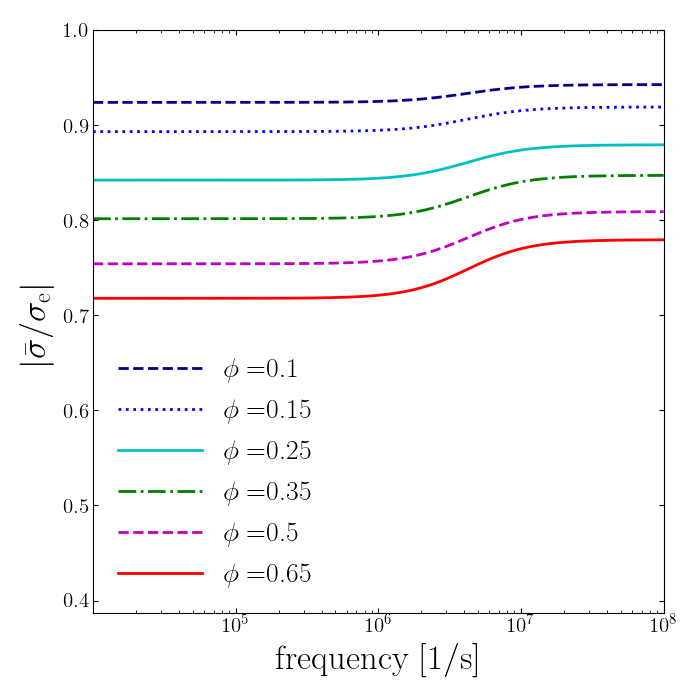}\label{fig:homogenization2}}
\subfigure[]{\includegraphics[height=0.4\linewidth]{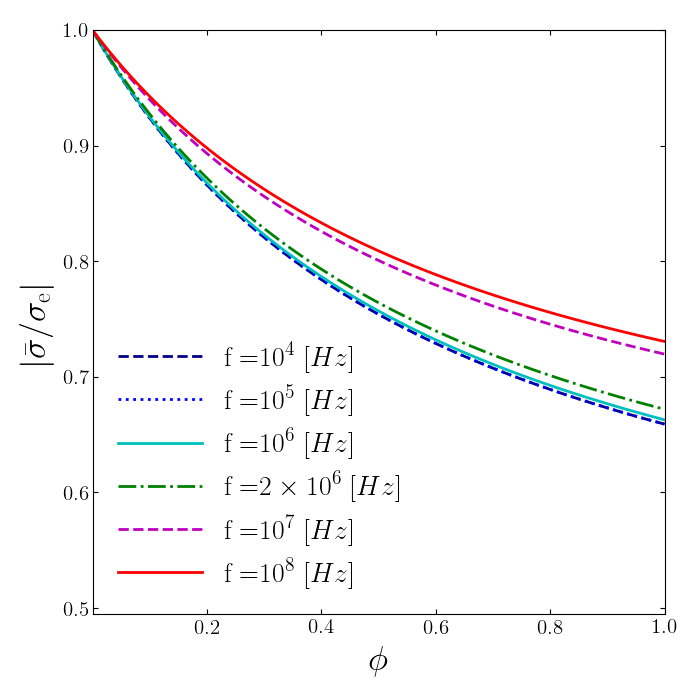}\label{fig:homogenization22}}
\caption{Equation \eqref{eq:homogenization2} for $\nu=0.46$ with $\sigma_e=1.3\ [S/m]$, $\sigma_c=0.6\ [S/m]$, and the cell radius set to $7\ [\mu m]$. (a,b) Low membrane conductance $S_L=1.9\ [S/m^2]$, (c,d) high membrane conductance $S_L=1.9\times 10^5 \ [S/m^2]$.}
\label{fig:homogenization}
\end{figure*}

On the other hand, one could identify relative complex permittivity $\epsilon^\ast=\epsilon' - j\epsilon''$ through its definition, \textit{i.e.}  the current $J$ is related to an alternating applied field $E$ via $J=\sigma_e E + j\omega\epsilon_0\epsilon^\ast E$. Comparing with equation \eqref{eq::mainCurrent}, we can write
\begin{align*}
&j\omega \epsilon_0 \epsilon^\ast \mathbf{E}_{ext} = (1 - \frac{\sigma_m}{\sigma_e}) n<\mathbf{P}> + n<\mathbf{S}>,\\
 \textrm{with}  \\
& \epsilon^\ast =\frac{\sigma_e}{j\omega \epsilon_0}\cdot ( (1 - \frac{\sigma_m}{\sigma_e}) \chi_p + \chi_s)=\epsilon' - j\epsilon'',
\end{align*}
where $\epsilon'$ is the dielectric constant, $\epsilon''$ is the loss factor of the material and $\epsilon_0=8.854\times 10^{-12}\ [F/m]$ is the dielectric permittivity of free space. Furthermore, the complex admittance is given by $Y^\ast=(\sigma_e + j\omega\epsilon_0\epsilon^\ast)A_{el}/H$. The admittance being the inverse of the impedance, $Z$, we have:
\begin{align}
 Z &= \frac{H}{A_{el}}\cdot \frac{1}{\sigma_e + j\omega \epsilon_0 \epsilon^\ast} \nonumber \\
 &=\frac{H}{ A_{el}} \cdot \frac{ \mathbf{E}_{ext}\cdot \mathbf{k}}{(\sigma_e \mathbf{E}_{ext} + (1 - \frac{\sigma_m}{\sigma_e}) n<\mathbf{P}> + n<\mathbf{S}>)\cdot\mathbf{k} } \label{eq::impedancefinal}
\end{align}
where $H$ is the distance between the electrodes and $A_{el}$ is the surface area of one electrode. We designate $Z_e=H/(\sigma_e A_{el})$ to define the dimensionless impedance as
\begin{align*}
\frac{Z}{Z_e}&= \frac{\sigma_e \mathbf{E}_{ext}\cdot \mathbf{k}}{(\sigma_e \mathbf{E}_{ext}+(1 - \frac{\sigma_m}{\sigma_e})n<\mathbf{P}> + n<\mathbf{S}>)\cdot\mathbf{k} }. \\
&=\frac{1}{1 + (1 - \frac{\sigma_m}{\sigma_e})\chi_p + \chi_s}
\end{align*}
Lastly, the complex conductivity $\sigma^\ast$ of the material is related to the admittance according to $Y^\ast=\sigma^\ast A_{el}/H=1/Z$ which implies that:
\begin{align*}
\sigma^\ast=\sigma_e\cdot (1 + (1 - \frac{\sigma_m}{\sigma_e}) \chi_p + \chi_s).
\end{align*}
Therefore, the values for admittance and the effective conductivity coincide.

\paragraph*{Effects of cellular pairwise interactions.}

Cell-cell interactions influence the effective conductivity of the aggregate, for example in the case of spherical particles Jeffrey (1973) \cite{jeffrey1973conduction} showed that pairwise interactions produce a correction term of order $\mathcal{O}(\phi^2)$ to the effective conductivity of a random dispersion of spherical particles. Importantly, Jeffrey used the \textit{twin spherical harmonics} invented by Ross (1968) \cite{ross1968potential} to account for two-particle interactions. This method was later applied to coated spheres by Lu and Song (1996) \cite{lu1996effective}. \cite{lu1996effective} derived the general expression for the effective conductivity of a random suspension of coated spheres, that is accurate up to order $\mathcal{O}(\phi^2)$:
\begin{align}
\frac{\bar{\sigma}}{\sigma_e}= 1 + 3\phi \theta_1 + \frac{3\phi^2 \theta_1^2}{1 - \phi \theta_1 } + \frac{K_2^\ast \phi^2}{1 - \phi \theta_1 },\label{eq:homogenization2}
\end{align}
where $3\phi\theta_1\equiv \chi_s + (1 - \frac{\sigma_m}{\sigma_e})\chi_p$ is the polarizability factor. $K_2^\ast$ accounts for higher order interactions due to detailed pair distribution of particles; here we neglect this last term as we are not considering detailed information about the microstructure of the aggregate (see Hasselman and Johnson (1987) \cite{hasselman1987effective} for a similar result). As shown by \cite{lu1996effective}, this estimate stays within Hashin-Shtrikman bounds \cite{hashin1962variational} up to high volume fractions of about $\phi\approx 0.6$. Figure \ref{fig:homogenization} illustrates the dependence of the effective conductivity on the volume fraction and on the frequency predicted by equation \eqref{eq:homogenization2} after setting $K_2^\ast=0$.

\section{Coarse-grained dynamics} \label{sec::III}
In the present modeling approach, cell-level dipole moments are the resolved observables that relate cellular properties to multicellular features. In this section, we introduce \textit{time-domain} governing equations for the dipole moments.

We consider a spherical cell of radius $R$ immersed in a mean electric flux $<J>$,
\begin{align}
C_m \frac{\partial [u]}{\partial t} + \bigg( S_m + \frac{(2+\phi)\sigma_e \sigma_c}{R \tilde{\sigma}}\bigg) [u] = \frac{3 \sigma_c\sigma_e}{\tilde{\sigma}}E \cos\theta, \label{eq::GEQ1}
\end{align}
and we seek a governing equation for the membrane dipole by integrating equation \eqref{eq::GEQ1} over cell membranes and multiplying by $\sigma_e$,
\begin{align*}
C_m \frac{d }{d t}\mathbf{P} + \bigg( S_m + \frac{(2+\phi)\sigma_e \sigma_c}{R \tilde{\sigma}}\bigg) \mathbf{P}= - \frac{\sigma_e^2 \sigma_c }{\tilde{\sigma}} A  \mathbf{E}. 
\end{align*}
Here we assumed a uniform conductance over the cell membranes. We divide both sides with the cell volume and obtain:
 \begin{align*}
C_m \frac{d}{d t} \mathbf{p} + \bigg( S_m + \frac{(2+\phi)\sigma_e \sigma_c}{R \tilde{\sigma}}\bigg) \mathbf{p}= -\frac{ 3\sigma_e^2 \sigma_c}{R\tilde{\sigma}} \mathbf{E} 
 \end{align*}
Due to this mean-field approximation, each cell evolves ostensibly independently from each other. Therefore, we define the coarse grained electrodynamics of an individual cell with
\begin{align}
 \dot{\mathbf{p}}&= -\gamma \mathbf{p} -  \alpha \mathbf{u}(t),  \label{eq::TD3}
\end{align} 
where we represent the time-dependent model for the electric flux with $\mathbf{u}(t)$ and define
\begin{align*}
&\alpha= \frac{3\sigma_e \sigma_c}{C_m R \tilde{\sigma}}  &\gamma = \frac{S_m}{C_m} +\frac{\sigma_e\sigma_c}{R C_m \tilde{\sigma}}(2+ \phi)
\end{align*}
and the stimulating field is that of the mean electric field in the matrix at any given time $t$,
\begin{align*}
 \mathbf{u}(t)&\equiv \sigma_e\mathbf{E}(t)=\sigma_e \kappa^{-1} \mathbf{E}_{ext}(t).
\end{align*}
Note that $\mathbf{u}(t)$ is the mean field approximation for the external electric current that each cell feels.

Starting from the microscopic equation \eqref{eq::TD3} we shall derive a mesoscopic model for an ensemble of cells. The basic idea is to view an ensemble of dipoles as random variables, and subsequently treat equation \eqref{eq::TD3} as a Langevin equation for the dynamical evolution of the random variables. In 1908, Langevin introduced the concept of equation of motion of a random variable \cite{langevin1908theorie} and through his formulation of the dynamical theory of Brownian motion, he initiated the subject of stochastic differential equations \cite{nelson1967dynamical}.  

To bridge the microscale to the mesoscale, we are interested to know the probability distribution of dipole moments in an aggregate. In stochastic systems (\textit{e.g.} in many condensed matter systems that are in contact with a heat bath) a successful strategy is to start from a Langevin equation describing the evolution of a single particle. Then, through appropriate averaging procedures, one arrives at a Fokker-Planck equation describing the evolution of the probability distribution of that particle. Eventually, the independence assumption implied in the mean field approximation allows to define the total probability distribution $W(\{p_k\},t)$ as the product of that of individual particles $W(\{p_k\},t)=\Pi_i W_i(p_i,t)$.

Unfortunately, this procedure fails in the system of cell aggregates due to the deterministic nature of the electrodynamics of cells. Through direct numerical simulations we know that the evolutionary trajectory of cell dipoles is not a stochastic process; in fact equation \eqref{eq::TD3} already suggests that polarizations are given by a deterministic equation. In the next subsection, we solve this problem by considering the randomness in cell parameters (see figure \ref{fig:alphagamma}) and devise a stochastic interpretation. We also mention the interesting work of Takayasu \textit{et al.} (1997) \cite{takayasu1997stable} who took a similar strategy to analyze the conditions for the emergence of power-law distributions in specific discrete Langevin models of the form $x(t+1) = b(t)x(t) + f(t)$ where both $b(t)$ and $f(t)$, are random variables.

\subsection{The indistinguishable stochastic replica}
\begin{figure}
\begin{center}
\subfigure{\includegraphics[width=\linewidth]{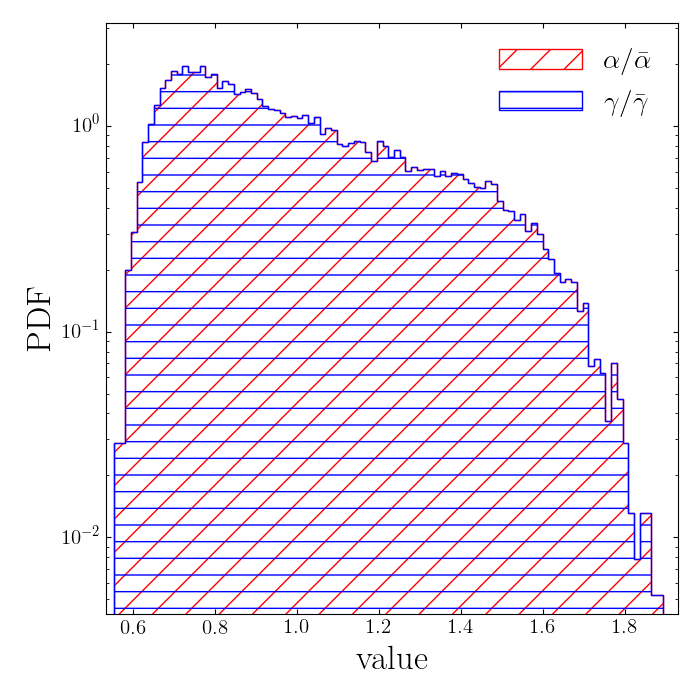} \label{subfig::alphagamma}} \quad \quad
\end{center}
\caption{Distribution of $\alpha$ and $\gamma$ parameters used in numerical simulation. The densities follow exponential profile in the middle range.}
\label{fig:alphagamma}
\end{figure}
The main source of randomness in the dielectric response of multicellular systems is the diversity of cell parameters, \textit{i.e.} namely $\alpha$ and $\gamma$. Here we exercise an alternative viewpoint to replace a diverse ensemble of deterministic cells with an indistinguishable ensemble of stochastic elements. In particular we consider a \textit{fiducial} random-walk process in cell parameters such that the empirical probability density of cells matches that of actual ones at any timestep. This is achieved by defining two random processes $A(t)$ and $B(t)$ for each cell such that
\begin{align*}
\alpha_i &=\bar{\alpha} +  A_i (t), \\
\gamma_i &= \bar{\gamma} + B_i (t),
\end{align*}
and
\begin{align*}
&\bar{\gamma}=<\gamma>, &\bar{\alpha}=<\alpha>.
\end{align*}
Figure \ref{fig:alphagamma} illustrates the distribution of these parameters. It is evident that around their mean values the distributions follow an exponential profile (note that the central part of this figure is linear and the y-axis is in logarithmic scale), however as a first step we approximate these random processes with Gaussian white noise
\begin{align*}
<B_i(t)>&=0\\
<A_i(t)>&=0\\
<A_i(t)A_j(t')> &=\alpha'^2 \delta_{ij}\delta(t-t')\\
<B_i(t) B_j(t')>&=\gamma'^2 \delta_{ij}\delta(t-t')\\
<A_i(t) B_j(t')>&=\epsilon\alpha'\gamma' \delta_{ij} \delta(t-t')
\end{align*}
where $\vert \epsilon\vert \le 1$ models the degree of cross-correlation between the two noise terms. Substituting in equation \ref{eq::TD3} yields
\begin{align}
 \dot{\mathbf{p}}_i&= -\bar{\gamma} \mathbf{p}_i -  \bar{\alpha}\mathbf{u}(t) - \mathbf{p}_iB_i(t) -  \mathbf{u}(t)A_i(t).  \label{eq::TD4}
\end{align}
Equation \eqref{eq::TD4} is a Langevin model with both additive and multiplicative noise terms. In analogy, the additive noise term models a \textit{heat bath} acting on the dipole and the multiplicative noise term models effects of a fluctuating barrier. Interestingly, such hybrid models of arithmetic and geometric Brownian motions have many applications in Physics \cite{steinbrecher2007extreme,schenzle1979multiplicative}, Biology \cite{may2008ecology} and finance \cite{shaw2015model}. 
 
We take equation \eqref{eq::TD4} along each spatial dimension as an independent stochastic differential equation in the random variable $X_t\equiv p_k$, then we combine the two Brownian motions into a single Brownian motion to arrive at the stochastic differential equation
\begin{align}
dX_t =& (-\bar{\alpha}u - \bar{\gamma} X_t)dt \nonumber \\
& + \sqrt{\alpha'^2u^2 + 2\epsilon \gamma' \alpha' u X_t + \gamma'^2 X_t^2}dW_t \label{eq::FSP}
\end{align}
where $W_t$ is a Wiener process. This is in fact a member of \textit{Pearson diffusions} that were extensively considered by Forman and S\o renson \cite{forman2008pearson}.

Here we emphasize that the replica stochastic aggregate must render a similar distribution of polarizations as in the actual system, therefore in general the processes $A(t)$ and $B(t)$ are not Markov processes with Delta autocorrelation in time. In fact the specificities of these random processes must be tuned to match the actual distribution. In the next section we simply consider white noise as a first approximation in this direction.

\subsection{The Fokker-Planck equation}
Equation \eqref{eq::FSP} constitutes the Stratonovich vector stochastic differential equations of a Langevin equation with a multiplicative noise term. The generic form of the coupled stochastic differential equation in terms of stochastic variables $\xi_i$'s reads
\begin{align*}
\frac{d\{\xi\}_i}{dt}=h_i(\xi_1,\xi_2,\xi_3,t) + g_{ij}(\xi_1,\xi_2,\xi_3,t)\lambda_j(t)
\end{align*}
with $g_{ij}\lambda_j$ are multiplicative noise terms that produce the noise-induced drift and diffusion components. $\boldsymbol{\lambda}$ has the following property,
\begin{align*}
&<\lambda_i(t)>=0, &<\lambda_i(t)\lambda_j(t+\tau)>=\delta_{ij}\delta(\tau)
\end{align*}
Then these equations can be treated as the starting point for deriving the corresponding Fokker-Planck equation. To this end, let $\xi_1=p_x$, $\xi_2=p_y$, and $\xi_3=p_z$ and $W(\xi_1,\xi_2,\xi_3,t)d\xi_1d\xi_2d\xi_3$ be the probability of finding a dipole in $d\xi_1d\xi_2d\xi_3$ at time $t$, then the Fokker-Planck equation reads
\begin{align*}
\frac{\partial W}{\partial t}=-\frac{\partial }{\partial \xi_i}(D_i W)+\frac{1}{2}\frac{\partial^2}{\partial \xi_i \partial \xi_j}(D_{ij}W)
\end{align*}
where the Einstein's summation rule is implied. By use of the Langevin equations, one can evaluate the statistical averages $<\cdot>$ in order to get the following set of equations for the drift and the diffusion coefficients \cite{Risken1996}:
\begin{align*}
D_i&=h_i(\{\xi\},t)+\frac{1}{2}g_{kj}(\{\xi\},t)\frac{\partial}{\partial \xi_k}g_{ij}(\{\xi\},t),\\
D_{ij}&=g_{ik}(\{\xi\},t)g_{jk}(\{\xi\},t).
\end{align*}

Hasegawa (2008) \cite{hasegawa2008stationary,hasegawa2008moment} considered solutions to the Fokker-Planck equations associated with Langevin equation \eqref{eq::TD4} in the Stratonovich stochastic calculus; furthermore, Mortensen (1979) \cite{mortensen1969mathematical} derived the Fokker-Planck equation associated to SDE \ref{eq::FSP} according to Ito stochastic calculus. Fortunately this system is separable and we can treat it one dimension at a time for our analysis, \textit{i.e.} the other two dimensions can be identically treated. The governing FPE for the variable $x\equiv p_k$ with the Stratonovich interpretation reads:
\begin{align}
\frac{\partial }{\partial t}&W(x,t)= \nonumber \\
&\frac{\partial}{\partial x}\bigg[\bar{\gamma}x+\bar{\alpha}u(t) - \frac{\chi}{2}(\gamma'^2x + \epsilon\gamma'\alpha' u(t))\bigg]W(x,t) \nonumber \\
+&\frac{1}{2}\frac{\partial^2}{\partial x^2} \bigg[\gamma'^2x^2 + 2\epsilon\gamma'\alpha' u(t) x + \alpha'^2u^2(t)\bigg]W(x,t) \label{eq::FPEF}
\end{align}
where $\chi=0,1$ corresponds to the Ito or the Stratonovich interpretations of the stochastic calculus respectively. 

\subsection{Analytical Treatment} 
The invariant probability distribution has a density that satisfies:
\begin{align}
\frac{d}{dx}W(x)=-\frac{(\bar{\gamma} + \frac{\chi}{2}\gamma'^2)x + (\bar{\alpha} + \frac{\chi}{2}\epsilon\alpha'\gamma')u}{\frac{1}{2}\gamma'^2 x^2 + \epsilon\alpha'\gamma'u x + \frac{1}{2}\alpha'^2 u^2}~W(x)\label{eq::inva}
\end{align}
Equation \eqref{eq::inva} resembles the invariant density of \textit{Pearson diffusion} processes that are characterized with a linear drift and quadratic diffusion coefficients. Therefore, we find that the Fokker-Planck equation \eqref{eq::FPEF} falls in the category of Pearson diffusion processes \cite{pearson1914tables}, whose stationary probability density is invariant under translation and scale transformations. Statistical properties of this class of models have been analyzed by Forman and S\o renson \cite{forman2008pearson} (for a brief summary see section 1.13.12 of \cite{iacus2009simulation}). Fundamentally, Pearson diffusions are viewed as the solutions to the following stochastic differential equation in the canonical parameterization:
\begin{align*}
dX_t = -\theta (X_t-\hat{\mu}) dt + \sqrt{2\theta (\mathtt{a} X_t^2 + \mathtt{b}X_t + \mathtt{c})} dW_t, 
\end{align*}
with $\theta>0$ being a scaling of time that determines how fast the distribution evolves, $\mathtt{a},\mathtt{b},\mathtt{c}$ are shape parameters such that the diffusion coefficient is well defined, and $W_t$ is a Wiener process. 

Pearson processes can lead to a variety of distributions depending on the parameters of the drift and the diffusion coefficients such as heavy or light tailed and symmetric or skewed profiles \cite{forman2008pearson}. We briefly report six basic subfamilies from \cite{forman2008pearson,iacus2009simulation} that are determined using criteria on the degree of the diffusion polynomial in the denominator of equation \eqref{eq::inva} denoted here by $\mathtt{deg}$, the sign of the leading coefficient (that in our case is strictly positive), and the discriminant $\Delta=\mathtt{b}^2 - 4\mathtt{a}\mathtt{c}$:
\begin{enumerate}
\item \textit{if $\mathtt{deg}=0$:} A Ornstein-Uhlenbeck process with a normal invariant density.
\item \textit{if $\mathtt{deg}=1$:} If $0<\hat{\mu}\le 1$ we obtain a Cox-Ingersoll-Ross process while for $\hat{\mu}>1$, we obtain a gamma invariant density.
\item \textit{if $\mathtt{deg}=2$ and $\Delta>0$ and $\mathtt{a}<0$:} A Jacobi diffusion with a Beta invariant density.
\item \textit{if $\mathtt{deg}=2$ and $\Delta>0$ and $\mathtt{a}>0$:} A Fisher-Snedecor process with a Fisher-Snedecor invariant density.
\item \textit{if $\mathtt{deg}=2$ and $\Delta=0$:} A Reciprocal gamma process with an inverse gamma invariant density.
\item \textit{if $\mathtt{deg}=2$ and $\Delta<0$:} for $\hat{\mu}\neq 0$ we obtain a Student diffusion with a skewed $t$ invariant density, while for $\hat{\mu}=0$ we obtain a scaled $t$-distribution.
\end{enumerate}
In the current case, and under a non-zero applied pulse, we can establish that the discriminant is always negative
\begin{align*}
\Delta = -4\big(\alpha'\gamma'u\big)^2(1-\epsilon^2)<0.
\end{align*}
Therefore, we expect that the probability density of the dipole moments along the z-axis (parallel to the applied pulse) is best described by a \textit{skewed Student} distribution (also known as Pearson type IV distribution). Even though in the transverse direction the mean electric pulse is negligible based on our mean field model, we can not totally neglect the influence of the electric fluctuations. To  first order approximation we treat the transverse direction by setting $\hat{\mu}=0$ while preserving the same diffusion term as in the parallel direction. This also ensures a symmetric probability density for positive or negative values. Therefore, we conclude that the distribution in the transverse direction must follow a scaled \textit{t}-distribution (also known as Pearson type VII distribution). 

\subsubsection{Stationary Probability Density}

\paragraph{In the direction parallel to the applied pulse.} We can directly integrate equation \eqref{eq::inva} to identify the invariant distribution
\begin{align}
W_s(x; \nu, c, a, \lambda)=  K \frac{ \exp\big\{2c \tan^{-1}\big(\frac{x-\lambda}{a} \big) \big\} }{\big(1 + \big(\frac{x-\lambda}{a} \big)^2 \big)^{\nu}}, \label{eq::station}
\end{align}
where
\begin{align*}
&a=\frac{\alpha'u}{\gamma'}\sqrt{1-\epsilon^2}, \quad \lambda=-\frac{\epsilon \alpha' u}{\gamma'}, \\
&\nu=\frac{\chi}{2}+\frac{\bar{\gamma}}{\gamma'^2}, \quad c=\frac{\epsilon\alpha'\bar{\gamma} - \bar{\alpha}\gamma'}{\alpha'\gamma'^2\sqrt{1-\epsilon^2}}, \\ \textrm{and} \nonumber \\
&K=\frac{\big\vert \Gamma (\nu + ic)\big\vert^2}{a\sqrt{\pi}\Gamma(\nu)\Gamma(\nu - 1/2)}.
\end{align*}
Equation \eqref{eq::station} provides the average and the variance of the polarizations in the stationary state (\textit{cf.} original work of \cite{pearson1895x}, and \cite{heinrich2004guide} for a useful guide):
\begin{align}
\mu &\equiv \mathbb{E}(X_t)= \frac{ac}{\nu -1} + \lambda =\frac{ \epsilon\alpha'\gamma'u - \bar{\alpha}u}{\bar{\gamma} - \gamma'^2}, \\
\sigma^2&\equiv \mathbb{E}(X_t^2)-\mathbb{E}(X_t)^2=\frac{a^2[(\nu - 1)^2 + c^2]}{(\nu-1)^2(2\nu -3)} \nonumber \\
&=\frac{\gamma'^2\mu^2+2\epsilon\gamma'\alpha' u\mu + \alpha'^2 u^2}{2(\bar{\gamma} - \gamma'^2)}, \label{eq::first2moments}
\end{align}
where we have let $\chi=1$ to obtain the last equalities. Equation \eqref{eq::station} can be independently verified by comparing it to model B of Hasegawa \cite{hasegawa2008moment} via the change of notations $\lambda_H\equiv \bar{\gamma}$, $I_H\equiv -\bar{\alpha}u$, $\alpha_H \equiv \gamma'$, $c_H\equiv c$, $b_H\equiv \nu$, $f_H\equiv-\lambda$ and $\beta_H \equiv \alpha'u$ (the $H$ subscript indicates Hasegawa's notation). Moreover, Hasegawa \cite{hasegawa2008moment} derived approximate equations for the evolution of the average and the variance in this model that we report here for completeness:
\begin{align}
\frac{d\mu}{dt}&=-\bar{\gamma}\mu(t) -\bar{\alpha}u + \gamma'^2\mu(t) + \epsilon\gamma'\alpha'u, \\
\frac{d\sigma^2}{dt}&=-2(\bar{\gamma}-\gamma'^2)\sigma(t)^2  + \gamma'^2\mu(t)^2 + 2\epsilon\gamma' \alpha' u\mu(t) + \alpha'^2u^2,  \label{eq::hasegawa}
\end{align}
which provide the analytical formula for the evolution of the probability density throughout the polarization process when replacing the static values $\nu$ and $c$ in equation \eqref{eq::station} by their dynamic counterparts:
\begin{align*}
\nu(t)&=\frac{a^2 + (\mu(t) - \lambda)^2 + 3\sigma(t)^2}{2\sigma(t)^2}, \\
c(t)&=\bigg(\frac{a^2+(\mu(t)-\lambda)^2 + \sigma(t)^2}{2a\sigma(t)^2}\bigg) (\mu(t)-\lambda).
\end{align*}
Figure \ref{fig::mom} gives a comparison between the results obtained with model \eqref{eq::hasegawa} and the direct numerical simulation of Mistani \etal \cite{mistani2019parallel}. A few remarks follow: (i) Hasegawa's moment equations are valid approximations up to order $\mathcal{O}((\delta x)^2)$ about the mean value of dipolar polarization per cell volume, and (ii) in the current model, the dynamics for $\mu(t)$ appears to be decoupled from $\sigma^2$, which is merely the result of assuming a linear dependence for the conductance term $F(x)\equiv -\bar{\gamma}x$, as well as a linear assumption for the multiplicative noise factor $G(x)\equiv x$. Breaking either of these assumptions introduces extra contributions from the variance to the mean dynamics. However, such modifications would alter the stationary distribution of polarizations, which the current model captures well. Therefore we expect that the observed discrepancies in the temporal evolution between the model prediction and the direct numerical simulation most likely stem from the nature of the noise term. This could be alleviated by introducing fractional-order temporal derivatives. Indeed, figure \ref{fig::mom} actually proves that the transient dynamics of dipolar moments does not simply follow an exponential function.

\paragraph{In the direction perpendicular to the applied pulse.} In the transverse direction, the stationary density follows a symmetric $t$-distribution, which is obtained by setting the skewness parameter to zero in the skewed $t$-distribution \eqref{eq::station}, \textit{i.e.} set $c=0$ and all previous equations are valid. This indicates that the following identity regulates the transverse diffusion,
\begin{align*}
\epsilon_\perp \frac{\alpha'_\perp}{\bar{\alpha}}=\frac{\gamma'_\perp}{\bar{\gamma}}
\end{align*}

\subsubsection{Statistical moments}
In the type IV Pearson diffusions considered in this work, for $n\ge 2$ the $n^{\rm{th}}$ statistical moment can be computed using the recurrence formula \cite{heinrich2004guide}:
\begin{align}
\mu_n = &\frac{a(n-1)}{(\nu-1)^2[2(\nu - 1) - (n-1)]}  \nonumber \\
&\times \bigg\{c(\nu - 1)\mu_{n-1} + a ((\nu - 1)^2 + c^2)  \mu_{n-2} \bigg\} \label{eq::recurr}
\end{align}
where by definition $\mu_0=1$ and $\mu_1=0$. It is straightforward to check the consistency between equations \eqref{eq::recurr} and \eqref{eq::first2moments}.

An important notion is the regime of existence for each of the moments. The first moment exists for $\nu>1$, otherwise $<x>=\pm \infty$. The variance exists for $\nu>3/2$, otherwise it increases arbitrarily fast. The third moment exists if $\nu>2$ and the fourth moment exists for $\nu>5/2$.
\begin{figure*}
\begin{center}
\subfigure{\includegraphics[width=0.45\linewidth]{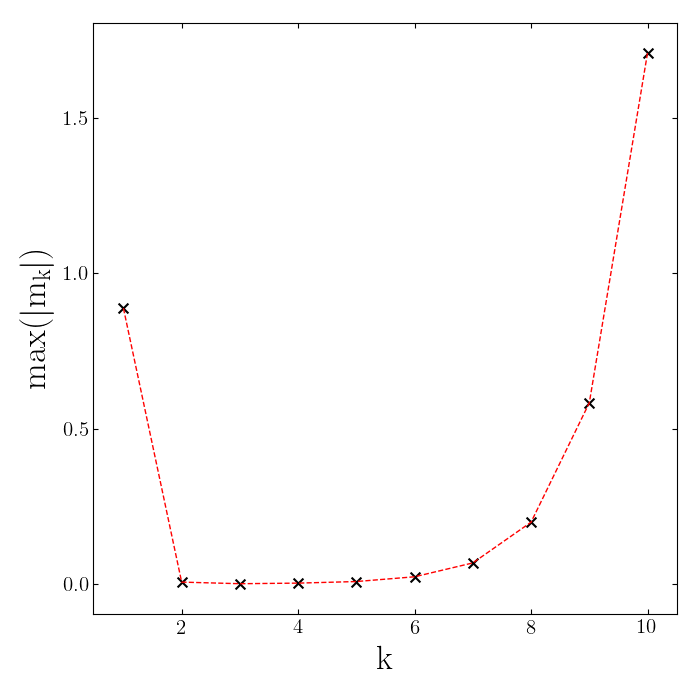} \label{subfig::maxmoment1}} \quad \quad
\subfigure{\includegraphics[width=0.45\linewidth]{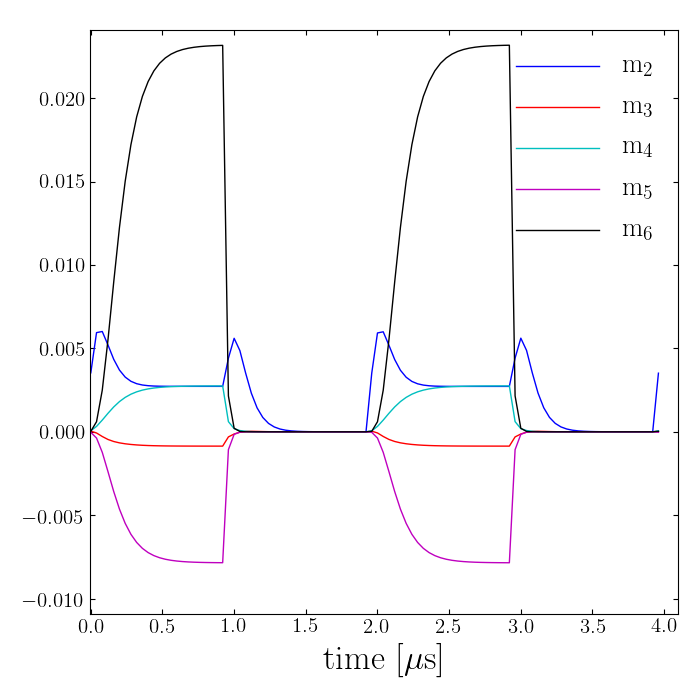} \label{subfig::v2}} \quad \quad
\end{center}
\caption{(Left) The maximum values of the statistical moments diverge for larger $k$ and (right) the evolution of the second to the sixth statistical moments of $\rm p_z\ [A/mm^2]$ in direct numerical simulations, \textit{i.e.} $m_k=<(p_z - <p_z>)^k>$. }
\label{fig:maxmoment}
\end{figure*}

\subsubsection{Fitting statistical moments}
\begin{figure}
\begin{center}
\subfigure{\includegraphics[width=\linewidth]{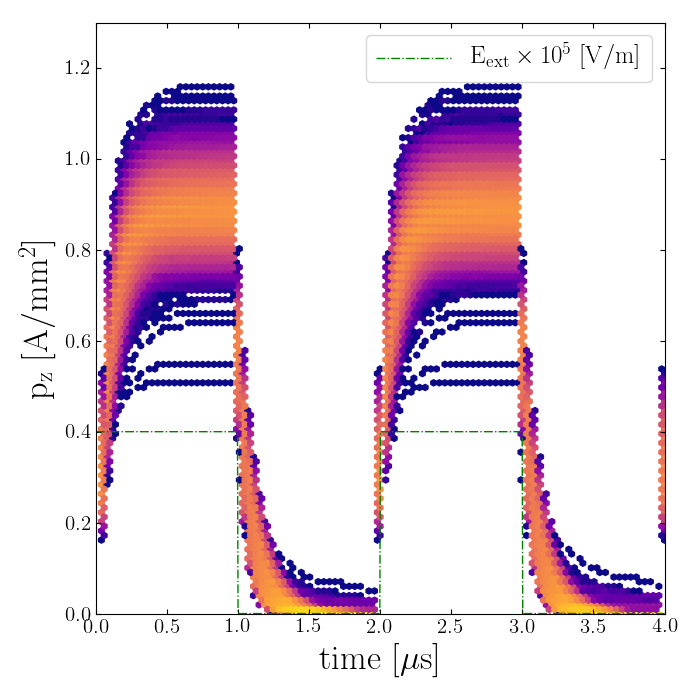} \label{subfig::sim}} \quad \quad
%\subfigure{\includegraphics[height=0.42\linewidth]{figs/model_density_evolve.png} \label{subfig::mod}} \quad \quad
\end{center}
\caption{Evolution of dipole moments using direct numerical simulation. }
\label{fig:compareSimMod}
\end{figure}
\begin{figure*}
\begin{center}
\subfigure{\includegraphics[height=0.42\linewidth]{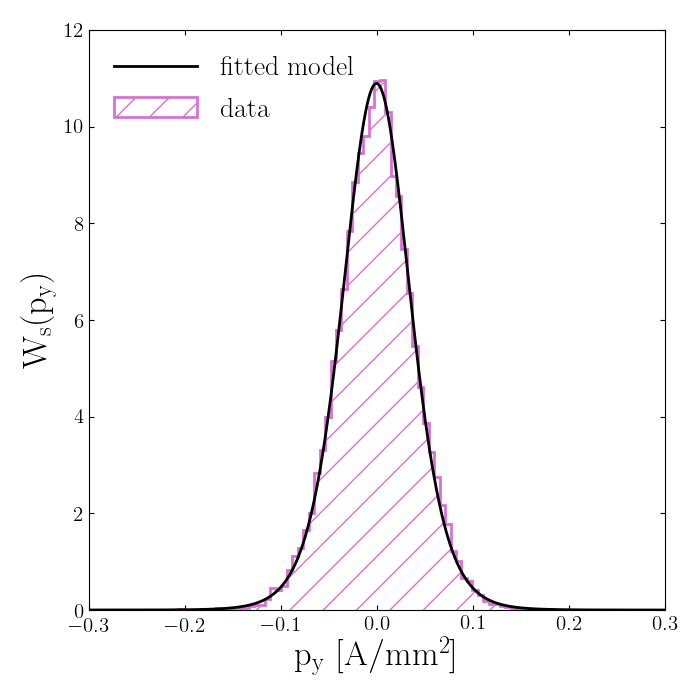} \label{subfig::fitty}} \quad \quad
\subfigure{\includegraphics[height=0.42\linewidth]{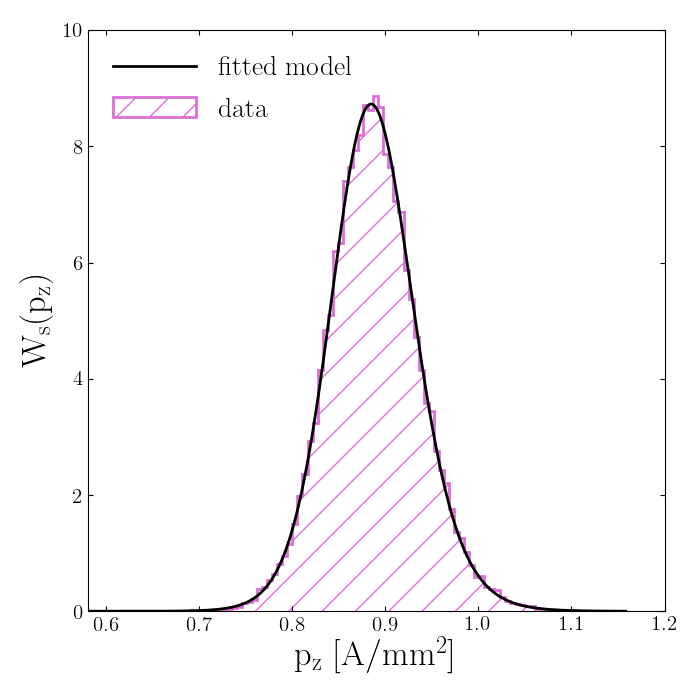} \label{subfig::fittz}} \quad \quad
\end{center}
\caption{The distribution of dipole moments in direct numerical simulation in comparison to model predictions. Measurements are made after $1\ [\mu s]$ of a step electric pulse (when stationarity is \textit{almost} achieved). Distributions follow (left) a symmetric $t$-distribution in the transverse direction (along the $y$-axis) with model parameters $(\nu, c, a, \lambda)\approx(5.054, 0.000, 0.108\ [A/mm^2], 0.000\ [A/mm^2])$, while (right) in the direction parallel to the applied pulse (along the $z$-axis), we observe a skewed $t$-distribution with model parameters $(\nu, c, a, \lambda)\approx (7.246, 0.888, 0.164\ [A/mm^2], -0.864\ [A/mm^2])$. Note that the $x$-axes is the absolute value of the polarization.}
\label{fig::fitts}
\end{figure*}
\begin{figure*}
\begin{center}
\subfigure{\includegraphics[width=\linewidth]{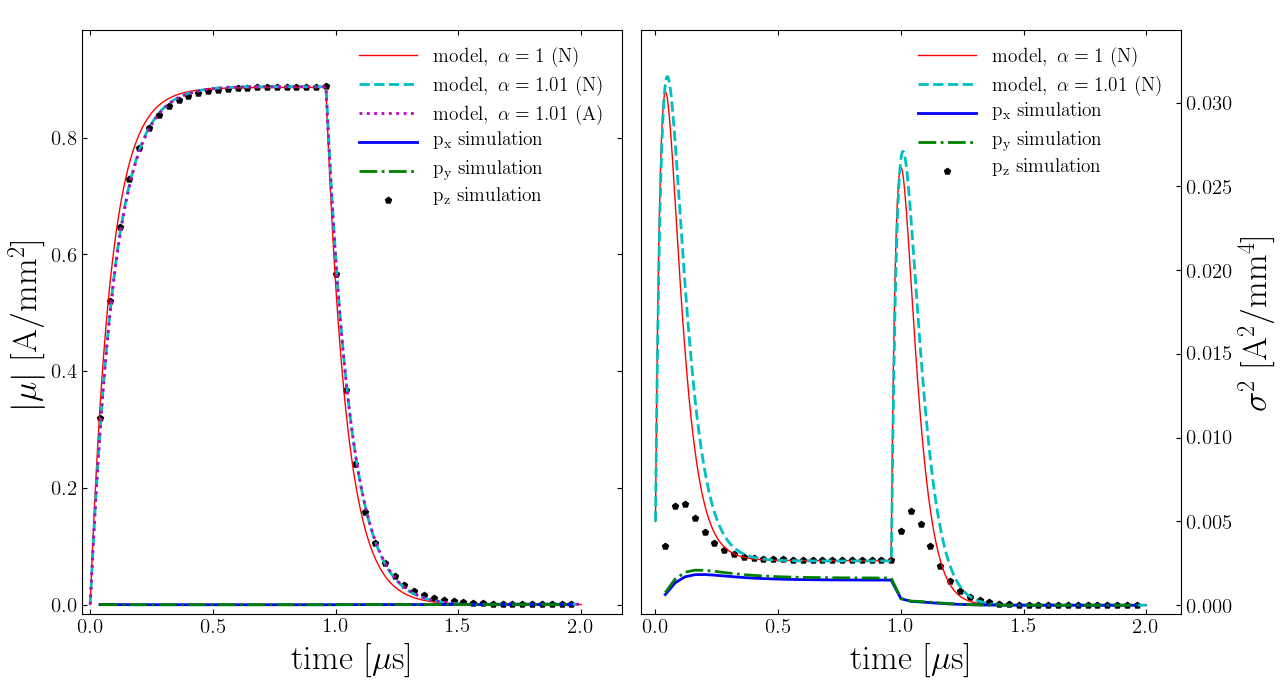} \label{subfig::mom1}} \quad \quad
\end{center}
\caption{The evolution of average and variance of dipole moments in direct numerical simulations. We chose $R=7\ [\mu m]$ and $\phi=0.13\times 0.00563$ to account for the free space surrounding the spherical tumor, \textit{i.e.} a box of $4\ [mm]$ on each side. The dotted magenta line in the left figure illustrates the analytical solution \eqref{eq::mu_fractional}. Letter N denotes numerical solution of FPE model while A indicates analytical solution of the FPE model.}
\label{fig::mom}
\end{figure*}
\begin{figure}
\begin{center}
\subfigure{\includegraphics[width=\linewidth]{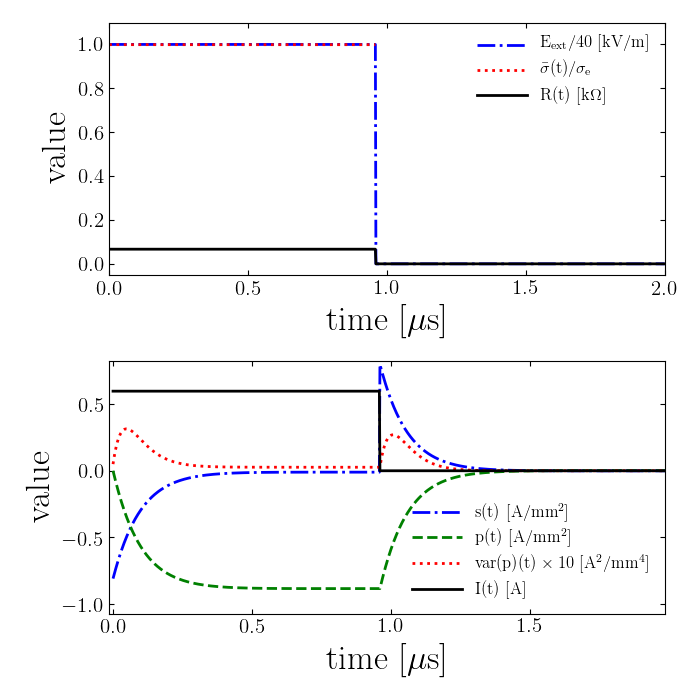} \label{subfig::mom2}} \quad \quad
\end{center}
\caption{Other time-domain properties based on our model for the case considered in figure \ref{fig::mom}. Immediately after switching off the applied pulse a reverse current is observed. }
\label{fig::mom2}
\end{figure}

We use the simple moment fitting approach introduced by Karl Pearson \cite{pearson1895x} (also see Heinrich's excellent guide \cite{heinrich2004guide}) to infer the four model parameters $(\nu, c, a, \lambda)$, which characterize the stationary probability density \eqref{eq::station}. Given simulation data, we can directly measure the first four statistical moments using the recurrence relation in equation \eqref{eq::recurr}, \textit{i.e.} we directly calculate $<x>$, $\mu_2$, $\mu_3$ and $\mu_4$. Afterwards, we can infer the unknown parameters using mean, variance and some intermediate quantities defined via the third and fourth moments:
\begin{align*}
\sqrt{\beta_1}&\equiv \frac{\mu_3}{\mu_2^{3/2}}=\frac{2c}{\nu-2} \sqrt{\frac{2\nu - 3}{(\nu - 1)^2 + c^2}}\\
\beta_2&\equiv \frac{\mu_4}{\mu_2^2} = \frac{3 (2\nu-3) [(\nu + 2)((\nu - 1)^2 + c^2) - 4(\nu - 1)^2]}{(\nu - 2)(2\nu - 5)}
\end{align*}
Thereafter, we compute the missing parameters according to
\begin{align*}
&\nu=\frac{5\beta_2 -6 \beta_1 - 9}{2\beta_2 - 3\beta_1 -6},\\
&c=\frac{(\nu - 1)(\nu-2) \sqrt{\beta_1}}{\sqrt{4(2\nu-3) - \beta_1(\nu-2)^2}},\\
&a=\sqrt{\mu_2 [(2\nu-3) - \frac{\beta_1}{4} (\nu - 2)^2] },\\
&\lambda=<x> - \frac{\sqrt{\mu_2\beta_1} (\nu-2)}{2}.
\end{align*}
Figure \ref{fig::fitts} illustrates the fitted model and its parameters. We observe that the model given in \eqref{eq::station} perfectly describes the results provided by our direct numerical simulation.
We compare the predictions of our model with the dynamics of the average and of the variance of dipole moments from our direct numerical simulations in figures \ref{fig::mom}--\ref{fig::mom2}. A few observations regarding the results of the direct numerical simulation can be drawn: (i) first, the decay of the average polarization does not follow an exponential decay; in fact it decays slightly slower than an exponential function, (ii) second, under a constant applied pulse, the variance increases initially but then decreases to reach a plateau, and after switching off the pulse, variance exhibits an uptick before decaying to zero. The observed uptick in the variance is a unique feature that is captured in our proposed model. 

We also solve for the model parameters in terms of the observed distribution parameters,
\begin{align*}
&\gamma'=\sqrt{\frac{\bar{\gamma}}{\nu - \frac{\chi}{2}}}, &\epsilon = \frac{1}{\sqrt{1 + \frac{a^2}{\lambda^2}}}, \\
&\alpha'=\frac{\bar{\alpha} \gamma'}{\epsilon \bar{\gamma} - c \gamma'^2 \sqrt{1-\epsilon^2}},    &u=-\frac{\lambda \gamma'}{\epsilon\alpha'}.
\end{align*}
The fitted values in figure \ref{fig::fitts} yield $\rm (\alpha', \gamma', \epsilon, u)=(2242.08\ [\sqrt{S/F}], 1446.9\ [\sqrt{S/F}], 0.983,, 0.568\ [A/mm^2])$, while $\bar{\gamma}=1.41\times 10^7\ [S/F]$ and $\bar{\alpha}=2.10\times 10^7\ [S/F]$. Figure \ref{fig::mom} depicts the comparison between our Fokker-Planck model and direct numerical simulations. We find that our model perfectly captures the qualitative trends observed in the distribution of the polarizations, while it is in good quantitative agreement. Also, it is important to develop numerical methods for the FPE in three spatial dimensions in order to faithfully compare these results, however the current one dimensional analytic treatment provides very encouraging results for the polarization component parallel to the applied pulse. We will investigate numerical solutions to the full FPE in future works.

\subsubsection{Fractional order evolution} \label{subsec::fractional}
In the case of Pearson diffusion, the statistical moments up to order $n<\mathtt{a}^{-1} +1$ exist. In particular, for $n\ge 2$, the autocorrelation decays exponentially \cite{bibby2005diffusion}:
\begin{align*}
corr(X_s, X_{s+t})=e^{-\theta t}. 
\end{align*}
In the current case, statistical moments exist up to order
\begin{align*}
n< 1 + 2\frac{\bar{\gamma}}{\gamma'^2}.
\end{align*}
Therefore, this analysis suggests that the autocorrelation of dipole moments is given by an exponential decay when
\begin{align*}
\gamma '^2 < 2 \bar{\gamma}.
\end{align*} 
This result provides a critical threshold for the diversity measure $\gamma_c' = \sqrt{2\bar{\gamma}}$ that regulates the autocorrelation function, \textit{i.e.} for $\gamma'<\gamma_c'$, we predict anomalous relaxation. Therefore, in our model, the observed anomalous relaxation in cell aggregate electroporation is associated with the diversity in the cellular structural parameters that is modeled through the parameter $\gamma$. In this case, the time derivative should be replaced with a fractional order derivative through $\dfrac{d}{dt}=\tau_1^{ \alpha-1}\dfrac{d^\alpha}{dt^\alpha}$, \textit{i.e.} $\tau_1$ is an arbitrary factor that has dimension of time. Here we assume $\tau_1=1$ and the governing equations read:
\begin{align}
\frac{d^\alpha\mu}{dt^\alpha}&=-\bar{\gamma}\mu(t) -\bar{\alpha}u + \gamma'^2\mu(t) + \epsilon\gamma'\alpha'u, \\
\frac{d^\alpha\sigma^2}{dt^\alpha}&=-2(\bar{\gamma}-\gamma'^2)\sigma(t)^2  + \gamma'^2\mu(t)^2 + 2\epsilon\gamma' \alpha' u\mu(t) + \alpha'^2u^2. \label{eq::Voigt}
\end{align}
In order to preserve the type of initial conditions appropriate in classical phenomena, \textit{i.e.} so that no extra initial conditions be needed, we adopt Caputo fractional derivative with $m-1 < \alpha \le m$ ($m$ is an integer number) \cite{caputo1967linear,gorenflo2008fractional}, which is defined by
\begin{align*}
_a ^C D_t^\alpha f(t) = \frac{1}{\Gamma (m-\alpha)} \int_{a}^t \frac{f^{(m)}(s)}{(t-s)^{\alpha-m+1}}~ds.
\end{align*}
Furthermore, defining fractional derivatives in the Caputo sense permits the application of the Laplace transform as a simple method of solution, see \cite{matlob2019concepts} for more details. In this case, the Laplace transform of Caputo derivatives reads:
\begin{align*}
    &\mathcal{L}[ _0 ^C D_t^\alpha f(t)]=s^\alpha f(s)-\sum_{k=0}^{m-1}f^{(k)}(0^+)s^{\alpha - k - 1}.
\end{align*}
Applying the Laplace transform to the set of equations \eqref{eq::Voigt} and assuming $\mu(0^+)=0$ and $\mu'(0^+)=0$ yields the transfer function $H(s)$ (or impulse response) for the average polarization:
\begin{align*}
\mu (s)=H(s) ~ u(s), \textrm{ with } H(s) =\frac{\epsilon\alpha'\gamma'- \bar{\alpha} }{s^\alpha + \bar{\gamma} - \gamma'^2}.
\end{align*}
Then the impulse response function is given by:
\begin{align*}
H(t)&= (\epsilon\alpha'\gamma'- \bar{\alpha}) t^{\alpha-1}E_{\alpha,\alpha}[ -(\bar{\gamma} - \gamma'^2) t^\alpha ],
\end{align*}
where $E_{\alpha,\alpha}$ is the Mittag-Leffler function that is generally defined as:
\begin{align*}
&E_{\alpha,\beta}[x]=\sum_{k=0}^\infty \frac{x^k}{\Gamma(k\alpha + \beta)}, &\alpha,\beta>0.
\end{align*}
Note that for $\alpha=\beta=1$, it is equivalent to the exponential function. Therefore the general solution to the average polarization is given by:
\begin{align*}
\frac{\mu(t)}{ \epsilon\alpha'\gamma'- \bar{\alpha} }&= \int_0^t \frac{E_{\alpha,\alpha}[ -(\bar{\gamma} - \gamma'^2) (t-\tau)^\alpha ]}{(t-\tau)^{1-\alpha}} u(\tau) d\tau.
\end{align*}
Particularly, the step response waveform (in response to $u(t)=u(0^+) \sum_{k=0}^\infty (-1)^k H(t-t_k)$, where $H(t)$ is the Heaviside function) reads:
\begin{align}
\mu(t) &= u(0^+) \frac{\epsilon\alpha'\gamma'- \bar{\alpha} }{\bar{\gamma} - \gamma'^2}  \nonumber \\
&\times \sum_{k=0}^\infty (-1)^k \bigg( 1 - E_{\alpha,1}[-(\bar{\gamma} - \gamma'^2)(t-t_k)^\alpha ] \bigg).  \label{eq::mu_fractional}
\end{align}
We found that a fractional order of $\alpha=1.01$ successfully describes the results obtained via direct numerical simulations, see figure \ref{fig::mom}. Alternatively, we could numerically evaluate equation \eqref{eq::Voigt} using a finite difference numerical scheme \cite{li2012finite,matlob2019concepts} which is basically to discretize Caputo derivative of order $0 < \alpha < 1$ using,
\begin{align*}
    _0 ^C D_t^\alpha f(t_{n+1}) &\approx \frac{(\Delta t)^{-\alpha}}{ \Gamma(2-\alpha)}\sum_{j=0}^n a_{j} \big( f_{n+1-j} - f_{n-j} \big),
\end{align*}
where $a_{j}=(j+1)^{1-\alpha} - j^{1-\alpha}$ and $f_j=f(t_j)$. Also, for $1<\alpha<2$ the discretization reads (\textit{c.f.} see equation 1.5 of Sun and Wu (2006) \cite{sun2006fully}):
\begin{align*}
_0 ^C D_t^\alpha f(t_{n+1}) \approx & \frac{ (\Delta t)^{-\alpha}}{\Gamma(3-\alpha)} \bigg[ f_{n+1}-f_n - b_{n-1} f'(0)\Delta t \\
& - \sum_{j=1}^{n-1} (b_{n-j-1} - b_{n-j} )(f_j - f_{j-1}) \bigg],
\end{align*}
where $b_j = (j+1)^{2-\alpha} - j^{2-\alpha}$ (see figure \ref{fig::mom} for comparison).

\section{Predictions \& Discussions} \label{sec::IV}
In this section we perform eight experiments based on the proposed model with the specifications given in tables \ref{tab:tb1}--\ref{tab:tb2}. In each experiment, we apply a Gaussian electric pulse given by:
\begin{align*}
E_{ext}(t)=E_0 \exp\bigg(-6\frac{(t-t_f/3)^2}{t_f^2} \bigg),
\end{align*}
where $E_0=40\ [kV/m]$ for a duration of $t_f=20\ [\mu s]$ to resolve electric response at smaller frequencies; note that the biggest frequency is determined by the maximum time-step size of integration, which we limit to $1\ [ns]$, while the smallest frequency is inversely proportional to duration of integration. Then impedance can be computed by equation \eqref{eq::impedancefinal}. 

We performed numerical integrations of the integer order system of ordinary differential equations \ref{eq::hasegawa} with the publicly available package \texttt{Scipy} \cite{2020SciPy-NMeth} with adaptive time stepping. In particular, we used \texttt{LSODA} algorithm, which automatically detects stiffness and switches between the non-stiff Adam and stiff BDF integration methods \cite{petzold1983automatic}. The fractional order ODEs \ref{eq::Voigt} are solved by implementing the discretization schemes discussed in subsection \ref{subsec::fractional}. The source code to solve the set of equations proposed in this manuscript and to reproduce the results of this section can be found at  \href{https://github.com/pourion/CAEP}{https://github.com/pourion/CAEP}.

\subsection{Time-domain response}
\begin{figure*}
\begin{center}
\subfigure[Configuration I.]{\includegraphics[width=0.45\linewidth]{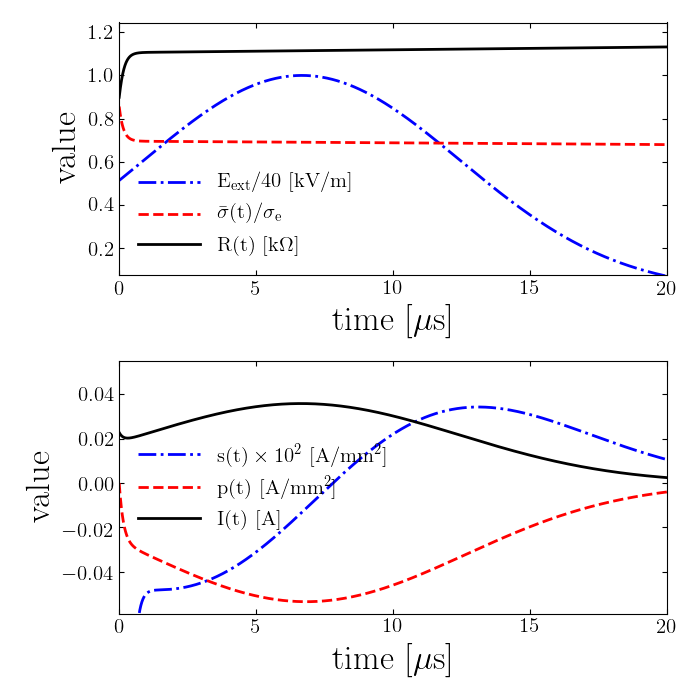} \label{subfig::TimeDom1}} \quad \quad
\subfigure[Configuration II.]{\includegraphics[width=0.45\linewidth]{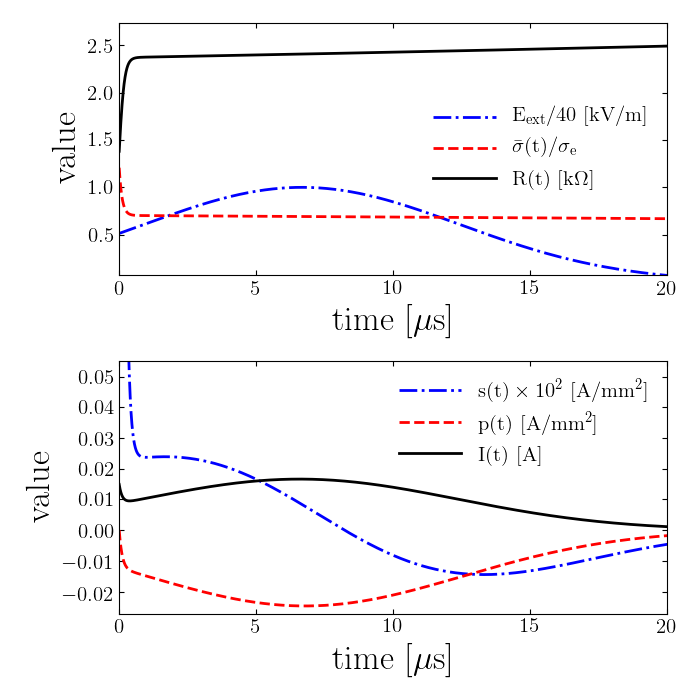} \label{subfig::TimeDom2}} \quad \quad
\subfigure[Configuration III.]{\includegraphics[width=0.45\linewidth]{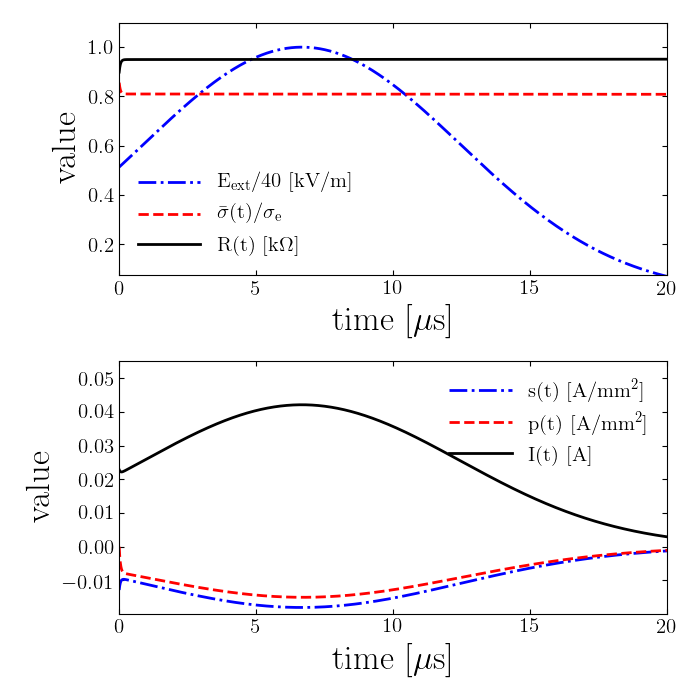} \label{subfig::TimeDom3}} \quad \quad
\subfigure[Configuration IV.]{\includegraphics[width=0.45\linewidth]{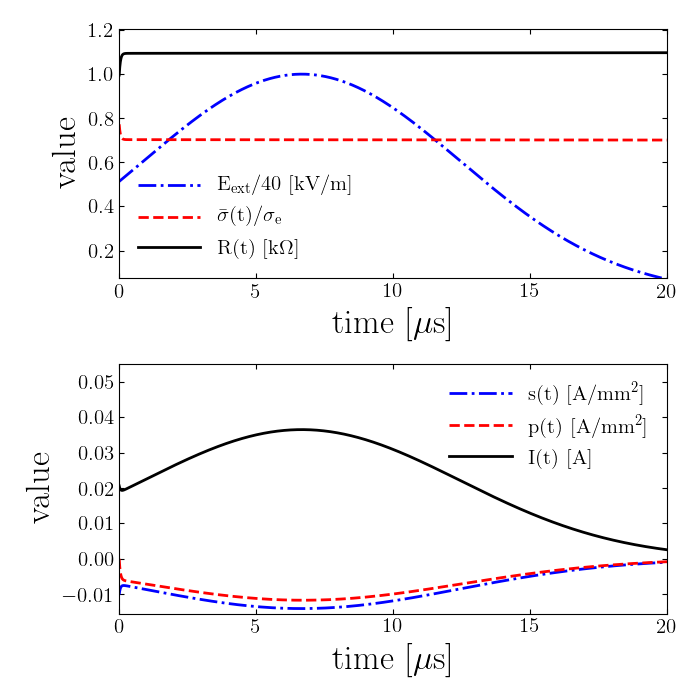} \label{subfig::TimeDom4}} \quad \quad
\end{center}
\caption{Experiments in the time domain.}
\label{fig::TimeDomainExps}
\end{figure*}

Figure \ref{fig::TimeDomainExps} illustrates the evolution of the instantaneous effective conductivity, current, resistance, and average value of the membrane and the cytoplasm polarizations during the application of the external Gaussian pulse. To make these predictions, we considered an integer order time derivative, \textit{i.e.} $\alpha=1$. 
\begin{table}
\caption{\label{tab:tb1} List of theoretical experiments based on the proposed framework. We consider a cell aggregate confined in a cubic box with side length $1\ [mm]$.}
\begin{ruledtabular}
\begin{tabular}{lccccc}
\textrm{\#} & \textrm{$ \sigma_c\  [S/m]$} & \textrm{$\sigma_e\ [S/m]$} & \textrm{$S_L\ [S/m^2]$} & \textrm{$\phi$} & \textrm{$\alpha$} \\
\colrule
I & $1.3$ & $0.6$  & $1.9$ & $0.3$ & $1$\\   
II & $0.6$ & $1.3$  & $1.9$ & $0.3$ & $1$\\  
III & $1.3$ &  $0.6$  & $1.9\times 10^5$ & $0.3$ & $1$\\  
IV & $1.3$ & $0.6$  & $1.9\times 10^5$ & $0.6$ & $1$
\end{tabular}
\end{ruledtabular}
\end{table}

At low membrane conductance (configurations I and II), we find that even though the magnitude of the cytoplasm polarization is negligible with respect to the membrane polarization, over time the cytoplasm dipole moment changes direction from anti-parallel to parallel with respect to the external field, which leads to a slowly increasing resistance felt at the electrodes. However, increasing the membrane conductance (configurations III and IV) has the effect of increasing the cytoplasm polarization at the expense of reducing the membrane polarization while both dipole moments remain anti-parallel to the external field. Increasing the membrane conductance enhances the overall current density and reduces the overall electric resistance of the aggregate.

We have shown how to compute the impedance directly from the time-domain FPE, which paves the way for more detailed studies of cell aggregates with nonlinear membrane processes such as the case of electroporation.

\subsection{Impedance spectroscopy} 
The purpose of this section is to understand the impedance as a function of frequency within cell aggregates. This analysis is important because it enables the resolution of the polarization processes and to relate them to their relaxation timescales, \textit{cf.} see the review by Asami (2002) \cite{asami2002characterization}. For example impedance spectroscopy is widely used as a technique to characterize ionic conductors, electroceramics, solid electrolytes, dielectric materials such as polymers and glasses as well as fuel cells and batteries \cite{macdonald2005impedance,orazem2008electrochemical,bonanos2002impedance}. 
\begin{table}
\caption{\label{tab:tb2} List of theoretical experiments based on the proposed framework. We consider a cell aggregate confined in a cubic box with side length $1\ [mm]$. In each case we chose $10$ different values for the varying parameter.}
\begin{ruledtabular}
\begin{tabular}{lccccc}
\textrm{\#} & \textrm{$ \sigma_c\  [S/m]$} & \textrm{$\sigma_e\ [S/m]$} & \textrm{$S_L\ [S/m^2]$} & \textrm{$\phi$} & \textrm{$\alpha$} \\
\colrule
V & $0.6$  & $1.3$ & $1.9$ & $[0.01, 0.8]$ & $1$\\   
VI & $1.0$  & $[0.5,1.5]$ & $1.9$ & $0.3$ & $1$\\
VII & $0.6$  & $1.3$ & $[1.9,1.9\times 10^5]$ & $0.3$ & $1$\\
VIII & $0.6$  & $1.3$ & $1.9$ & $0.3$ & $[0.9,1.1]$
\end{tabular}
\end{ruledtabular}
\end{table}
\begin{figure*}
\begin{center}
\subfigure[Configuration V - varying the volume fraction. ]{\includegraphics[width=0.45\linewidth]{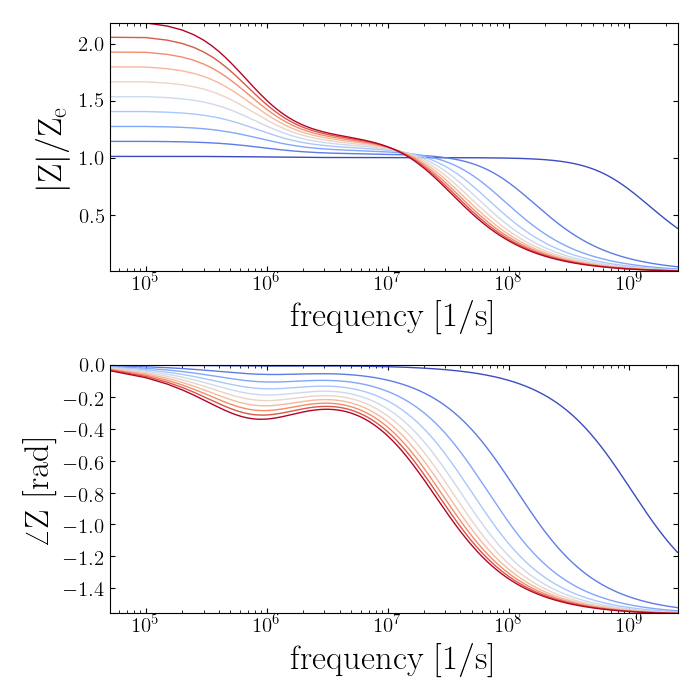} \label{subfig::predict1}} \quad \quad
\subfigure[Configuration V - varying the volume fraction.]{\includegraphics[width=0.45\linewidth]{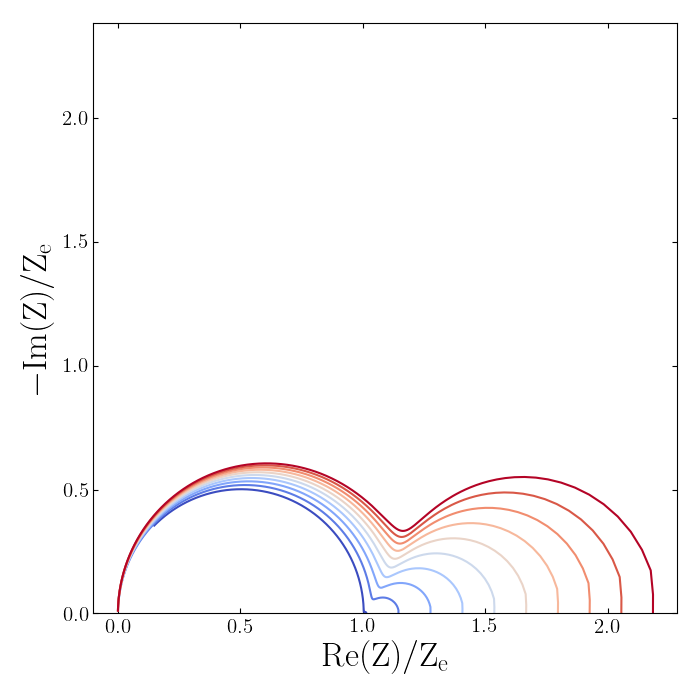} \label{subfig::predict12}} \quad \quad
\subfigure[Configuration VI - varying the extra-cellular matrix conductivity. ]{\includegraphics[width=0.45\linewidth]{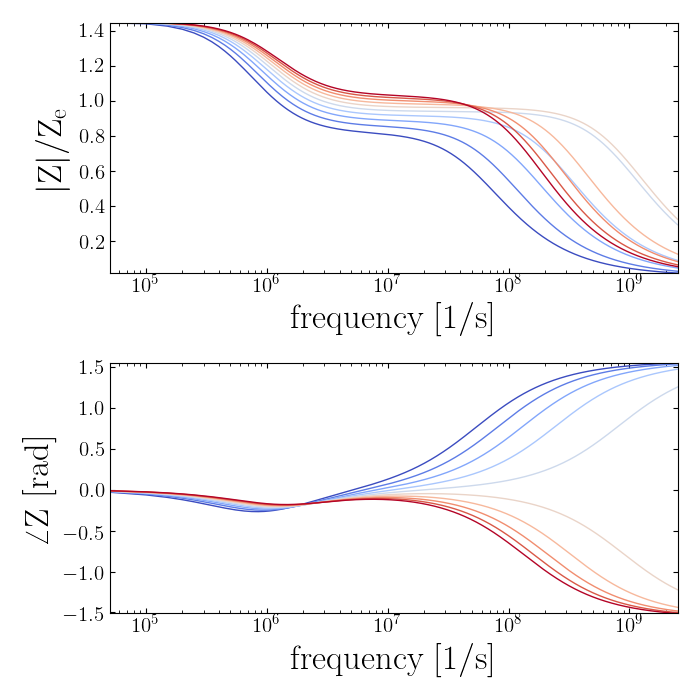} \label{subfig::predict2}} \quad \quad
\subfigure[ Configuration VI - varying the extra-cellular matrix conductivity.]{\includegraphics[width=0.45\linewidth]{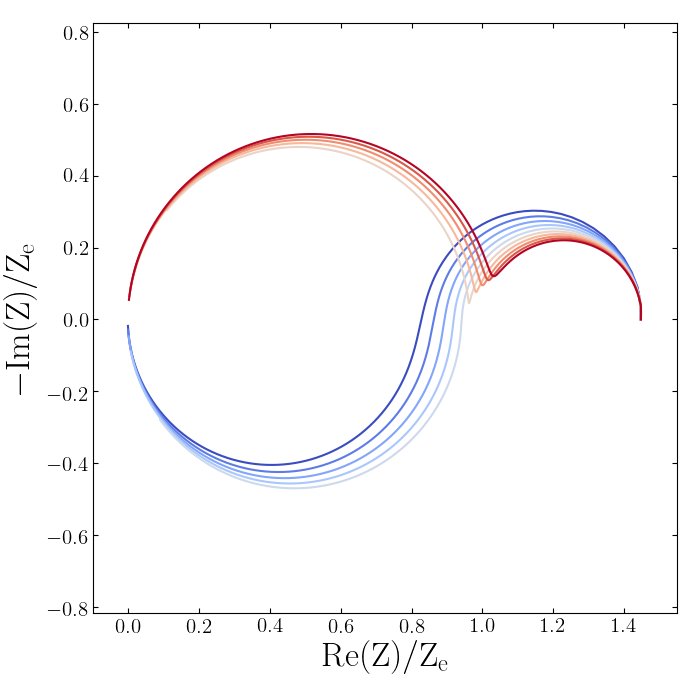} \label{subfig::predict22}} \quad \quad
\end{center}
\caption{Experiments in the frequency domain, in all figures warmer colors indicate higher values. Figure (a,b) show the effect of increasing volume fraction from $\phi=0.01$ to $\phi=0.8$. Figures (c,d) illustrate effects of increasing matrix conductivity in the range $\sigma_e=0.5\ [S/m]$ to $\sigma_e=1.5\ [S/m]$  while cytoplasm conductivity is fixed at $\sigma_c=1\ [S/m]$. Red colors correspond to the case of $\sigma_e>\sigma_c$ while blue colors correspond to $\sigma_e<\sigma_c$. }
\label{fig::predict1}
\end{figure*}
\begin{figure*}
\begin{center}
\subfigure[Configuration VII - varying the membrane conductance. ]{\includegraphics[width=0.45\linewidth]{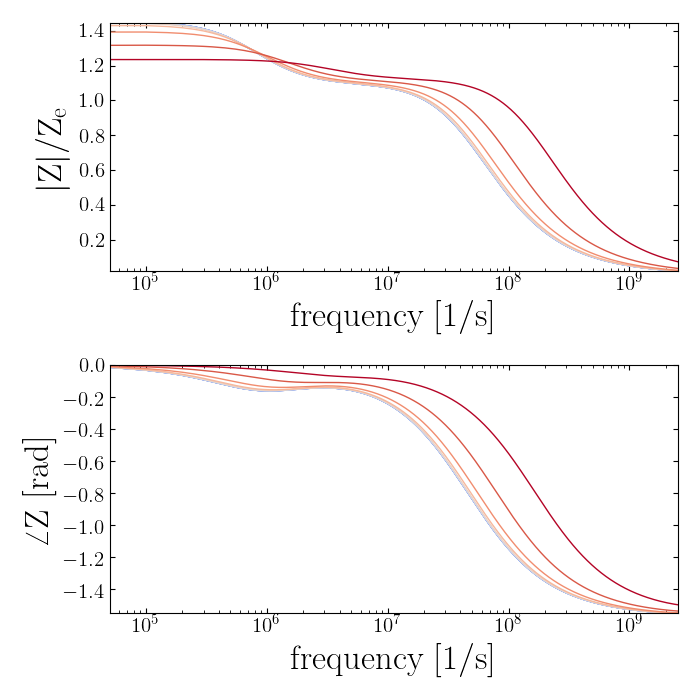} \label{subfig::predict3}} \quad \quad
\subfigure[ Configuration VII - varying the membrane conductance.]{\includegraphics[width=0.45\linewidth]{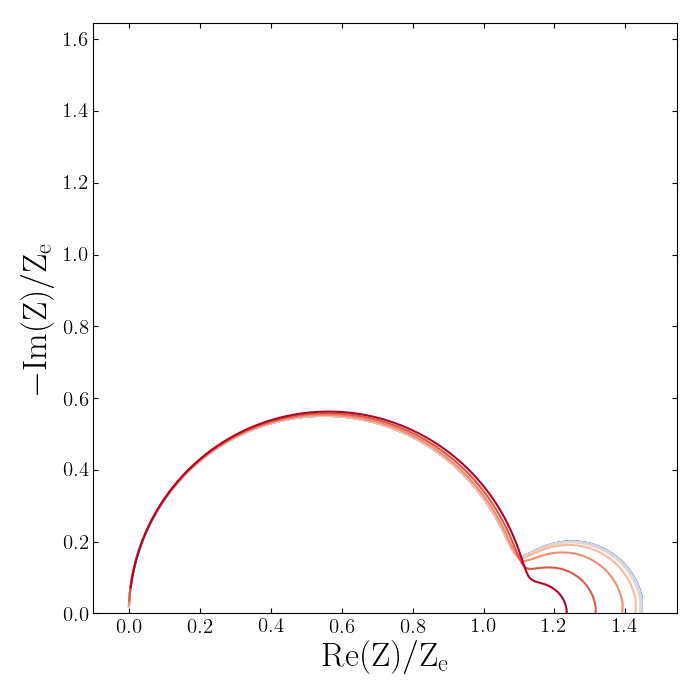} \label{subfig::predict32}} \quad \quad
\subfigure[Configuration VIII - varying the fractional order. ]{\includegraphics[width=0.45\linewidth]{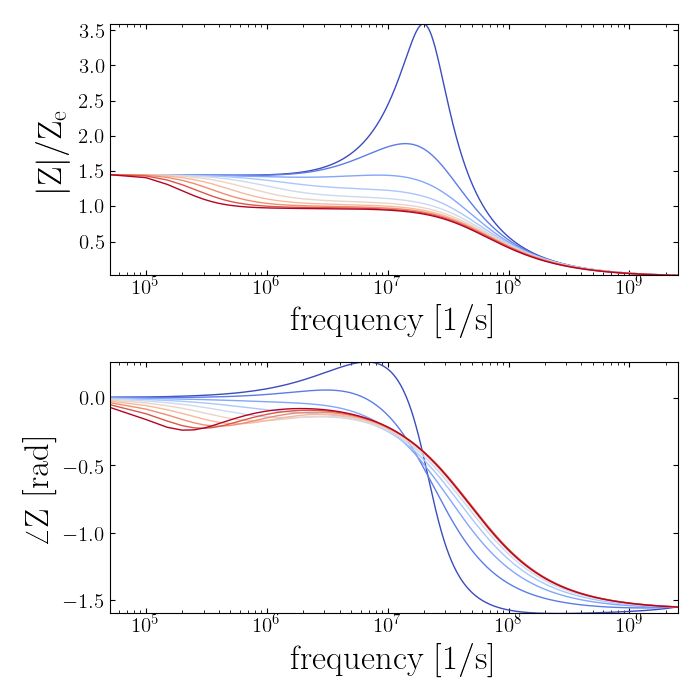} \label{subfig::predict4}} \quad \quad
\subfigure[ Configuration VIII - varying the fractional order.]{\includegraphics[width=0.45\linewidth]{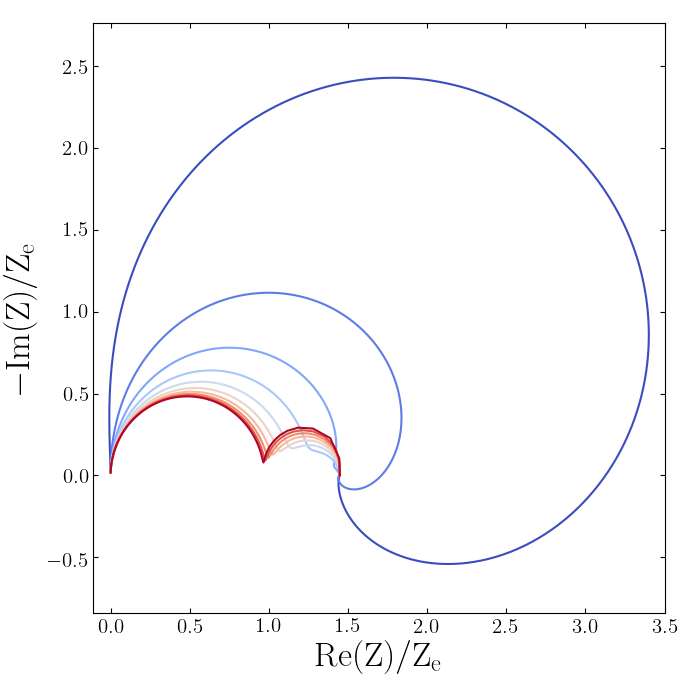} \label{subfig::predict42}} \quad \quad
\end{center}
\caption{Experiments in the time domain, in all figures warmer colors correspond to higher values. Figures (a,b) capture the effect of increasing the membrane conductance from $S_L=1.9\ [S/m^2]$ (bluer colors) to $S_L=1.9\times 10^5\ [S/m^2]$ (redder colors). Figures (c,d) illustrate the effects of increasing the fractional order from $\alpha=0.9$ to $\alpha=1.1$. Note that the curves for $\alpha<1$ are shown in blue, while for $\alpha>1$ red colors are used.}
\label{fig::predict2}
\end{figure*}

Figures \ref{fig::predict1}--\ref{fig::predict2} show the Bode and Cole diagrams calculated by equation \eqref{eq::impedancefinal} for $4$ different configurations and illustrate the effects caused by varying the volume fraction, the matrix conductivity, the membrane conductance and the order of the fractional derivative. 

In figures \ref{subfig::predict1}--\ref{subfig::predict12}, we gradually increase the volume fraction from $\phi=1\%$ to $80\%$ that increases the impedance at lower frequencies and reduces it at higher frequencies. More importantly, increasing the volume fraction appears to amplify a low-frequency semi-circle in the Cole diagram originating from cell membranes. 

In figures \ref{subfig::predict2}--\ref{subfig::predict22}, we change the matrix conductivity while keeping the other parameters fixed. We find that when the matrix is more conductive than the cytoplasm, the dielectric response of the cytoplasm lags behind that of the applied pulse. However for a cytoplasm more conductive than the matrix, we find that the cytoplam dielectric response leads the applied pulse. The latter behavior resembles the dielectric response of an inductive element that appears at high frequencies. 

In figures \ref{subfig::predict3}--\ref{subfig::predict32}, we gradually increase the membrane conductance and find that the semi-circle arc at low frequency gradually shrinks. The characteristic behavior of the present model is that the membrane determines the low frequency arc while the cytoplasm determines the high frequency arc.

In figures \ref{subfig::predict4}--\ref{subfig::predict42}, we vary the order of the fractional derivative. For $0 < \alpha < 1$, we observe a \textit{low frequency hook} effect, where an apparently inductive loop appears at low frequencies, \textit{i.e.} where the imaginary part of impedance becomes positive. In particular, we observe that our model predicts that, by increasing $\alpha$ towards $1$, the low frequency hook gradually shrinks, and the low frequency semi-circle becomes depressed. Interestingly, Cole and Baker (1941) \cite{cole1941longitudinal} reported an inductive response in their experiments with squid axons. Cole and Baker argued that inductive effects originated from the membrane of axons, which they modeled by an equivalent circuit composed of a resistor in series with an inductor that are connected in parallel with a capacitor. For a detailed discussion on the possible origins of inductive hooks we refer to Klotz (2019) \cite{klotz2019negative}. In fact, low frequency inductive impedance is ubiquitously found in impedance spectroscopy experiments with various systems such as Lithium ion batteries \cite{brandstatter2016myth}, proton exchange membrane fuel cells \cite{roy2007interpretation}, organic light emitting diodes (LEDs) \cite{bisquert2006negative}, Perovskite solar cells \cite{ghahremanirad2017inductive}, thin films on conductive substrates \cite{taibl2016impedance}, and corrosion of Chromium \cite{dobbelaar1990impedance}. It is well known that tissue impedance follows a depressed Cole (1940) equation, 
\begin{align*}
Z(\omega)&=R +\frac{ R_0 - R_\infty}{1 + (j\omega/\omega_0)^\alpha},
\end{align*}
where $\omega_0$ is the angular turnover frequency and $\alpha$ is a dimensionless number between zero and one \cite{cole1941dispersion, mcadams1995tissue, smye2007modelling}. It is generally established that it is the diversity of relaxation timescales that is responsible for the observed anomalous electric response of tissue environments \cite{zorn2002logarithmic}, which is the source of fractional order evolution in our model as well.

\section{Conclusion} \label{sec::V}
We have developed a theoretical framework based on a dipole decomposition of cell polarization into two parts: the membrane polarization, and the cytoplasm polarization. Based on this decomposition, we were able to evaluate effective properties of the aggregate environment such as effective conductivity and impedance. We also derived a time-domain governing Fokker-Planck equation that explains distributions of cellular polarizations in different volume fractions and at different frequencies. We showed that the effects of cell interactions can be easily included in the model. Our theory is generally applicable to triphasic structures that are ubiquitously found in nature, for example in modeling suspensions of biological cells and subcellular organella such as yeasts \cite{sugiura1964dielectric,asami1976dielectric}, E. coli \cite{fricke1956dielectric}, synaptosomes \cite{irimajiri1975dielectric}, and mitochondria \cite{pauly1960electrical}. The current work can be extended in several ways:
\begin{itemize}
\item In plants and micro-organisms, cells are covered by a cell wall that adds another layer to the dielectric structure; for details see Carstensen (1960) \cite{carstensen1968passive}. Hanai \textit{et al.} \cite{hanai1986theory,hanai1988number} showed the number of interfaces corresponds to the number of relaxations in the dielectric response of a heterogeneous system, which could explain how diversity in the dielectric properties of cells leads to anomalous relaxation. Therefore an extension of the current theory for multishell structures would be to develop $N$-phase interfacial polarization theories. 

\item Coupling the bulk relaxation processes in tissue environments, such as counterion polarization effects, with the interfacial polarization. In particular it was argued \cite{grosse1987permittivity} that counterion polarization effects contribute to the observed anomalous relaxation; thus it will be useful to examine such influences on the distribution of induced transmembrane potentials.

\item Under strong electric fields, nonlinear cellular phenomena occur. A well known example is the membrane breakdown that occurs under transmembrane potentials of about $\rm V_{ep}=0.2 \ [V]$, in a process referred to as electroporation \cite{neumann1972permeability,kotnik2019membrane}. Other phenomena include mechanical effects such as the alignment of non-spherical cells with an applied field, or the swelling effects due to water uptake caused by an increase in the membrane permeability.
 
\item Another interesting extension would be to consider the effects of gap junctions on the induced transmembrane potentials. Gap junctions are electrical connections between neighboring cells that provide direct pathways for ion transport in multicellular systems. Gap junctions are key regulators for embryonic development due to their ability to regulate transmembrane voltages; therefore understanding their interplay with an external electric stimulation poses new opportunities to control embryonic development and a new pathway to understand and control patterning in biological organisms.
\end{itemize}

\begin{acknowledgments}
Part of this research was funded by ARO W911NF-16-1-0136. The direct numerical simulations performed in this work used the Extreme Science and Engineering Discovery Environment (XSEDE), which is supported
by National Science Foundation grant number ACI-1053575 and ASC-150002. The authors acknowledge the Texas
Advanced Computing Center (TACC) at The University of Texas at Austin for providing HPC and
visualization resources that have contributed to the research results reported within this paper.
\end{acknowledgments}

\appendix

\bibliography{references}% Produces the bibliography via BibTeX.

\end{document}